\documentstyle[12pt]{article}

\include{epsf} 
\begin{document}

\title{\Large \bf Dynamical model and nonextensive statistical mechanics of
a market index on large time windows}

\author{ \large \bf M. Ausloos$^1$ and K. Ivanova$^2$ \\ $^1$ GRASP and SUPRAS,
B5, Sart Tilman, \\ B-4000 Li\`ege, Belgium\\$^2$ Pennsylvania State 
University,
\\ University Park, PA 16802, USA \\ }



\maketitle

\begin{abstract}

The shape and tails of partial distribution functions (PDF) for a financial
signal, i.e. the S\&P500 and the turbulent nature of the markets are linked
through a model encompassing Tsallis nonextensive statistics and leading to
evolution equations of the Langevin and Fokker-Planck type. A model originally
proposed to describe the intermittent behavior of turbulent flows describes the
behavior of normalized log-returns for such a financial market index, for small
and large time windows, both for small and large log-returns. These turbulent
market volatility (of normalized log-returns) distributions can be sufficiently
well fitted with a $\chi^2$-distribution. The transition between the small time
scale  model of nonextensive, intermittent process and the large 
scale  Gaussian
extensive homogeneous fluctuation picture is found to be at $ca.$ a 
200 day time
lag. The intermittency exponent ($\kappa$) in the framework of the Kolmogorov
log-normal model is found to be related to the scaling exponent of the PDF
moments, -thereby giving weight to the model. The large value of 
$\kappa$ points
to a large number of cascades in the turbulent process. The first Kramers-Moyal
coefficient in the Fokker-Planck equation is almost equal to zero, indicating
''no restoring force''. A comparison is made between normalized log-returns and
mere price increments.

\end{abstract}

\vskip 2cm


\vskip 2cm


\vskip 2cm

\newpage

\section{Introduction}

The time lag dependent price increments, returns, log-returns, normalized
log-returns of financial market indices, stocks and foreign currency exchange
markets are known to be non-gaussian distributions and rather exhibit 
fat-tailed
power-law distributions \cite{fama,stanleybook,stanley_nat,3,gopi}. 
The origin of
the so called large volatility characterized by such fat tailed 
distributions is
a key question; the fat tails in such data are thought to be caused by  some
''dynamical process'' through a hierarchical cascade of short and long-range
volatility correlations, though Gopikrishnan et al. consider that correlations
and tails have different origins \cite{gopi}. Destroying all correlations, e.g.
by shuffling the order of the fluctuations, is known to cause the fat tails
almost to vanish. It is still an open question whether both the fat-tailed
power-law of partial distribution functions (PDF) of the various 
volatilities and
their $evolution$ for {\it different time delays} in financial markets can be
described.

The fat tails indicate an unexpected high probability of large price changes.
These extreme events are of utmost importance for risk analysis. They are
considered to be a set of strong bursts in the energy dissipation of so called
clusters of high price volatility. In so doing the PDF and the fat tail event
existence are thought to be similar to the notion of intermittency in turbulent
flows \cite{neto}. Indeed employing the Fokker-Planck equation approach
\cite{risken} recent studies \cite{friedrich1,friedrich2,MichaJohnopt,tokyo2}
have shown that the dynamics of a market results from a flow of information
between long and short time scales. Since the distributions of returns obey a
Fokker-Planck equation, the time evolution of the price $increment$ signal
($\Delta y$) measured for a time lag $\Delta t$ is governed by a Langevin
equation \cite{risken}

\begin{equation} \frac{d \Delta y}{d\Delta t} = D^{(1)}(\Delta 
y(\Delta t),\Delta
t) + \eta(\Delta t) \sqrt {{D^{(2)} (\Delta y(\Delta t),\Delta t)}}, \label{le}
\end{equation} the drift $D^{(1)}$ and diffusion $D^{(2)}$ coefficients being
those of the Fokker-Planck equation
\cite{friedrich1,friedrich2,MichaJohnopt,tokyo2}. It is often assumed that
$\eta(\Delta t)$ is a correlated noise with gaussian statistics. Thus such a
dynamics may be analogous to the dissipation of energy from large to small
spatial scales in three dimensional turbulence as pointed out already in
\cite{neto,ghash,schmitt}.

On the other hand, the non-Gaussian character of the fully developed turbulence
\cite{frisch} has been linked to nonextensive statistical physics
\cite{tsallis,tsallis_bukman,tsallis_wilk,tsallis_arim,tsallis_beck1,
tsallis_beck2,tsallis_beck3,Sattin1,Sattin2}. Whence recently there has been a
large number of studies, e.g.
\cite{neto,tsallis_wilk,michael,borland,gupta_tsallis,kozuki} of financial
markets employing the nonextensive statistics including those involving fully
developed turbulence approach as in \cite{frisch}. The 
nonextensivity, i.e. some
anomalous scaling of classically extensive properties like the 
entropy, is linked
to a single parameter $q$, e.g. in the Tsallis formulation of nonextensive
thermostatistics.

In this paper on the study of the behavior of a financial index, i.e. the
S\&P500, on $large$ time windows, we apply as in \cite{michael,kozuki} for
$short$ time windows, a recently suggested model of hydrodynamic 
turbulence that
serves as a dynamic foundation for nonextensive statistics
\cite{tsallis_beck1,tsallis_beck2,tsallis_beck3}. Indeed long time lag effects
must be also investigated. Furthermore, it is known that some 
distinction must be
made concerning the type of financial market $signal$ which is 
examined. We will
compare results based on price increment and normalized log-return\footnote{\it
throughout this report the natural logarithm will be always used.} time series.

In Sect. 2, we describe the distribution of returns for the daily closing price
signal of the S\&P500 index for the time interval between Jan. 01, 
1980 and Dec.
31, 1999, thus a series of $N=$5056 data points. Daily closing price values of
the S\&P500 index for the period of interest were downloaded from 
Yahoo web site
\footnote{\it http://finance.yahoo.com}. We characterize the tail(s) of the
distribution for various $\Delta t$'s, i.e. from 1 up to 40 days, and will
observe the value of the PDF tails, for such time lags, outside the 
best gaussian
through the data. In Sect. 3, we calculate the power law exponents 
characterizing
the $integrated$ distribution of the (normalized log-) returns over different
time lags for the S\&P500 index daily closing price through a detrended
fluctuation analysis and a power spectral density analysis point of 
view. Results
are compared to shuffled data for estimating the value of the error bars. In
Sect. 4, Tsallis statistical approach is outlined, and distributions of
(normalized log-) returns for time lags between $\Delta t$ = 1 to 40 days are
examined. It is found that the $q$-value of the nonextensive entropy 
converges to
a value = 1.22 for $\Delta t$ = 40~days, starting with $q$ = 1.39 for 
$\Delta t$
= 1, values similar to those reported for the intraday evolution of other
financial indices, e.g. NASDAQ in \cite{tsallis_et_al} and slightly lower than
those for S\&P500 minute data \cite{michael}. The probability density 
$f_{\Delta
t}(\beta)$ of the volatility $\beta$ in terms of the standard deviation of the
normalized log returns of the S\&P500 for different time lags is found to obey
the $\chi^2$-distribution. The intermittency exponent ($\kappa$) of the
Kolmogorov log-normal model is found to be related to the scaling 
exponent of the
PDF moments, -thereby giving weight to the model. The large value of $\kappa$
points to a large number of cascades in the underlying turbulent process.

In Sect. 5, the usual Fokker-Planck approach for treating the time-dependent
probability distribution functions is summarized and coefficients 
governing both
the Fokker-Planck equation for the distribution function of normalized log
returns and the Langevin equation for the time evolution of normalized log
returns of daily closing price signal of S\&P500 are obtained. We will notice
that there is ''no restoring force''.

Therefore we present for the first time a coherent theory linking the shape and
tails of partial distribution functions for long and short time lags 
of the daily
closing values of a financial signal and connect the often suggested turbulent
nature of the markets to a model encompassing nonextensive statistics and
evolution equations of the Langevin and Fokker-Planck type. We are  aware that
the number of data points of the time series ($N=$5056) might seem quite small
with respect to other studies involving millions of data points.  Some previous
work had indicated the possible use of 5000 or so data point series in order to
obtain scaling arguments and ingredients for models \cite{4,5}. Clearly the
relative error bars or confidence interval being  roughly proportional  to
$N^{1/2}$ have to be taken with caution. Thus the conclusions universal value
might be debated upon. Nevertheless one positive aspect might be that  scaling
effects are not {\it too sensitive} to $N$. We do warn however throughout the
report that some caution might also be taken concerning the stationarity of the
data. These $caveat$ might only be resolved through further work.

\section{Distribution of returns}

There are several ways of calculating the returns in a financial 
market. A simple
one represents price increment $\Delta y(t,\Delta t)$ or difference between the
value of the price signal $y(t)$ at time $t+\Delta t$ and its value 
at time $t$.
Log-returns, i.e. the logarithm of the price ratio $\tilde y(t)=ln[y(t+\Delta
t)/y(t)]$ are also used sometimes. Below we consider the {\it normalized
log-returns} $Z(t,\Delta t)=(\tilde y(t) - <\tilde y>_{\Delta 
t})/\sigma_{\Delta
t}$, where $<\tilde y>_{\Delta t}$ denotes the average and $\sigma_{\Delta t}$
the standard deviation of $\tilde y(t)$ for a given $\Delta t$. The normalized
log-returns $Z(t,\Delta t)$ depend on the time $t$ and the time lag $\Delta t$.
However, in order to simplify the notations and whenever possible 
without leading
to confusion and misunderstanding we will drop the explicit writing of one or
both variables. Daily closing price values of the S\&P500 index for the period
between Jan. 01, 1980 and Dec. 31, 1999 will serve as a standard 
financial signal
$y(t)$, thus $N=$5056 data points.

The distribution of the normalized log-returns $Z(t,\Delta t)$ of the daily
closing price signal for S\&P500 index for the period between Jan. 01, 1980 and
Dec. 31, 1999, for $\Delta t$ = 1 day are plotted in Fig. 1. A fit is first
attempted with a Gaussian distribution for $small$ values of the 
increments, i.e.
the central part of the distribution. This central part distribution is well
fitted with such a Gaussian type curve within the interval $Z\in[-3,3]$ but
departs from the Gaussian form outside this interval. The negative and positive
tails of the distribution outside the Gaussian curve are found both to be equal
to -3.1. In the case $\Delta t\ge$ 1~day, it is observed that the best Gaussian
range is the same as for $\Delta t=1$~day (Fig. 2) but the outside tail values,
as estimated for the various $\Delta t$'s of interest, are slightly different,
and found to decrease with $\Delta t$; they are reported in Table 1. These
findings ($\Delta t$-independent Gaussian range and tail exponent behavior) are
at odd with the expectation that the PDF tends toward a Gaussian for large
$\Delta t$. Some Bayesian-like analysis of the PDF's, i.e. allowing for the
expected Gaussian width behavior, has been done, with the appropriate 
conclusion.
However, the error bar on the various widths do not allow for a statistically
convincing evidence through the comparison of variance classical test. Same for
the tail exponents which are obtained from a very small number of data points.
Therefore it seems appropriate to pursue further the PDF analysis through other
techniques that one allowing to extract a PDF tail from the 
difference between a
raw data histogram and a central region Gaussian fit.

Notice that the  Oct 19, 1987 and Oct. 27, 1997 crashes, as studied elsewhere
\cite{sornettecrash87,auslooscrash97} are represented by isolated dots at $ Z=
-23$ and  and $Z=- 7 $ respectively\footnote{\it Although the absolute value of
the S\&P500 drop in price is of order of 60 units in both crashes, since the
price has increased and due to the nonlinearity of the logarithm the 
value of $Z$
is much smaller at the crash in 1997.}. The value of $Z= -8.7$ represents the
aftershock crash of the Oct. 26, 1987 \cite{auslooscrash97}. The analysis
presented in the present study is not designed to capture such 
extreme events nor
their effects.

\section{Time correlations}

There are different estimators for the long and/or short range dependence of
fluctuations correlations \cite{taqqu}. In all cases it is useful to test the
null hypothesis  debated \cite{3,gopi}  whether the fat tails are related to
or/and caused by long-range volatility correlations. Destroying all 
correlations
by shuffling the order of the fluctuations, is known to cause the fat tails
almost to vanish. A Kolmogorov-Smirnov test (not shown) on shuffled data has
indicated us the statistical validity of the numerical values and the
statistically acceptable meaning of the displayed error bars.

Through the (linearly)  detrended fluctuation analysis (DFA) method, see e.g.
\cite{hu}, we show first that the {\it long range correlations} of 
daily closing
price signal of S\&P500 for the time interval of interest, are 
brownian-like. The
method has been used previously to identify whether long range 
correlations exist
in non-stationary signals, in many research fields such as e.g. finance
\cite{4,5}, bioinformatics \cite{smkw}, cardiac dynamics 
\cite{plamennature1996}
and meteorology \cite{buda_dfa,kimaeeta,bunde}. The DFA concepts are therefore
not repeated here. For an extensive list of references see \cite{hu}. Briefly,
the signal time series $y(t)$ is $first$ $integrated$, to ''mimic'' a 
random walk
$Y(t)$. The time axis is next divided into $k+1$ non-overlapping boxes of equal
size size $n$; one looks thereafter for the best (linear) trend, 
$z(n)$, in each
box, and calculates the root mean square deviation of the (integrated) signal
with respect to $z(n)$ in each box. The average of such values is 
taken at fixed
box size $n$ in order to obtain

\begin{equation} F(n) = \sqrt{\left<{1 \over n } {\sum_{i=kn+1}^{(k+1)n}
{\left[Y(i)-z(i)\right]}^2}\right>}. \end{equation}

\noindent The box size is next varied over the $n$ value. The 
resulting function
is expected to behave like $ n^{1+H_{DFA}}$ indicating a scaling law
characterized by a (Hurst) exponent $H_{DFA}$. For the (integrated) 
daily closing
price signal of the S\&P500 index, a scaling exponent 
$1+H_{DFA}=1.52\pm0.01$ is
found (Fig. 3) in a scaling range extending from about 1 week to 
about 250~days,
i.e. 1 year.

Along the same line of thoughts the scaling properties of the normalized
log-returns $Z(t,\Delta t)=(\tilde y(t) - <\tilde y>_{\Delta t})/ 
\sigma_{\Delta
t}$ have also been tested for different time lag values, i.e. $\Delta 
t=1, 5, 10,
15, 20,$ $25, 30, 35, 40$~days (Fig. 4). The DFA function, as defined 
here above,
of the integrated normalized log returns for time lag 1~days behaves as {\it
white noise} and has a Hausdorff measure equal to zero (later see its power
spectrum in Fig. 6). However non trivial scaling properties occur for 
the series
of normalized log returns as soon as $\Delta t\ge 1$~day. The values of the
scaling exponents and the maximum box size $n_x$ (in days) for which 
the scaling
holds for each DFA-function are given in Table \ref{table2}, while the
DFA-functions together with fitting lines are plotted in Fig. 4. The values of
the Hausdorff measure of the normalized log-returns signals varies with $\Delta
t$ from $H_{DFA}=0.27\pm0.04$ for $\Delta t=5$~days to 
$H_{DFA}=0.43\pm0.01$ for
$\Delta t=40$~days. Recall that $H_{DFA}= 0.5$ corresponds to Brownian motion.
The value of the maximum box size for which the scaling holds $n_x$ 
is related to
the periodicities of the normalized log-returns signals defined by the value of
the time lag as $n_x\approx3.5\Delta t$. The data and the power law fit of this
functional dependence are plotted in the inset of Fig. 4. The value 
of the slope
~1.0 is the same as the one found by Hu et. al. \cite{hu} when studying the
effects of sinusoidal trends and noise on the (so called second order in
\cite{hu}) DFA technique.

The power spectrum of the daily closing price signal of S\&P500 $S(f)\sim
f^{-\mu}$ with spectral exponents $\mu_1=2.41\pm 0.06$ and $\mu_2=1.95\pm 0.03$
with a scale break at 250~days is shown in Fig. 5. The scaling 
properties of the
power spectrum of the shuffled daily closing price signal of S\&P500, in which
e.g. the amplitudes are randomly shuffled are shown in the inset of 
Fig. 5. Such
a scaling spectral exponent $\mu=0$ is the signature of a white noise like
behavior. Recall that $\mu =2.0$ corresponds to Brownian motion.

We have also checked for scaling behavior and possible periodicities 
in the power
spectrum of the time series of the normalized log returns $Z(t,\Delta 
t)=(\tilde
y(t) - <\tilde y>_{\Delta t})/\sigma_{\Delta t}$ for different 
(selected) values
of the time lag $\Delta t=1,5,20,35,$ $40$~days (Fig. 6). A white noise like
behavior of the power spectrum of such returns  always occurs for $1/f$ $\le$
$\Delta t $~days; e.g., dashed line in Fig. 6 for $f<1/128$ and $\Delta
t=40$~days. This is in accordance with the results of the DFA analysis (Fig. 4
and Table \ref{table2}). A scaling behavior is found at large frequencies $f$
satisfying the relationship $\mu=2H_{DFA}+1$, as indicated  in Fig. 6, e.g. by
the dashed line with slope $\mu=1.86$, for $f>1/128$ for  the case  $\Delta
t=40$~days.

Periodicities in the power spectrum of the normalized log return time 
series for
$\Delta t>1$~day were expected to be found since these periods are somewhat
embedded into the time series by the way they are obtained and the Fourier
transform technique. It is easily observed that the maxima and the 
minima of the
spectrum correspond to harmonics and subharmonics of $1/\Delta t$.

\section{Tsallis statistics}

Based on the scaling properties of multifractals \cite{mf}, Tsallis
\cite{tsallis,tsallis_1995} proposed a generalized Boltzmann-Gibbs
thermo-statistics through the introduction of a family of non-extensive entropy
functional ${\cal S}_q$ given by:

\begin{equation} {\cal S}_q=k \frac{1}{q-1}\left(1-\int p(x,t)^q dx\right),
\label{ts} \end{equation} with a single parameter $q$ and where $k$ is a
normalization constant. The main ingredient in Eq.(\ref{ts}) is the
time-dependent probability distribution $p(x,t)$ of the stochastic 
variable $x$.
The functional is reduced to the classical extensive Boltzmann-Gibbs 
form in the
limit of $q\longrightarrow 1$. The Tsallis parameter $q$ characterizes the
non-extensivity of the entropy. Subject to certain constraints the 
functional in
Eq.(\ref{ts}) seems to yield a probability distribution function of the form
\cite{neto,tsallis,tsallis_beck1,michael,kozuki}

\begin{equation} p(x) = \frac{1}{Z_q}\left\{1+\frac{C \beta_0
2\alpha(q-1)|x|^{2\alpha}} {2\alpha-(q-1)}\right\}^{-\frac{1}{(q-1)}}
\label{tsallis} \end{equation} for the stochastic variable $x$, where

\begin{equation} \frac{1}{Z_q}=\alpha\left\{\frac{C \beta_0 2\alpha(q-1)}
{2\alpha-(q-1)}\right\}^{1/2\alpha} \,
\frac{\Gamma\left(\frac{1}{q-1}\right)}{\Gamma\left(\frac{1}{2\alpha}\right)
\Gamma\left(\frac{1}{q-1}-\frac{1}{2\alpha}\right)} \label{zq} 
\end{equation} in
which $C$ is a constant and $0<\alpha\le 1$ is the power law exponent of the
potential $U(x)=C|x|^{2\alpha}$ that provides the ''restoring force'' $F(x)$ in
Beck model of turbulence
\cite{tsallis_beck1,tsallis_beck2,tsallis_beck3,Sattin2}. The latter 
is described
by a Langevin equation

\begin{equation} \frac{dx}{dt}=-\gamma F(x)+R(t) \label{force} \end{equation}
where $\gamma$ is a parameter and $R(t)$ is a gaussian white noise. A non-zero
value of $\gamma$ corresponds to providing energy to (or draining from) the
system by the outside \cite{Sattingrangas} The parameter $\beta_0$ in
Eq.(\ref{tsallis}) and (\ref{zq}) is the mean of the fluctuating volatility
$\beta$, i.e. the local standard deviation of $|x|$ over a certain 
window of size
$m$ \cite{neto}. We will use this model assuming that the normalized 
log returns
$Z(t,\Delta t)$ represent $the$ stochastic variable $x$, as in 
Eq.(6), or $\Delta
y$ in Eq.(1). We will search whether Eq.(\ref{tsallis}) is obeyed for $x\equiv
Z(t,\Delta t)$, thus studying $p(x)\equiv p_{\Delta t}(Z)$ for 
various time lags
$\Delta t$.

Just as in Beck model of turbulence\footnote{The approach used here 
was recently
suggested to be an appropriate model of hydrodynamic turbulence for financial
markets in \cite{kozuki}.} \cite{tsallis_beck1,tsallis_beck2,tsallis_beck3} we
assume that the volatility $\beta$ is $\chi^2$-distributed with 
degree $\nu$ (see
another formula in \cite{Sattin2}):

\begin{equation} f_{\Delta t}(\beta) \equiv
\frac{1}{\Gamma(\nu/2)}\left(\frac{\nu}{2\beta_0} \right)^{\nu/2}
\beta^{\nu/2-1}\exp\left(-\frac{\nu\beta}{2\beta_0}\right), \quad \nu > 2,
\label{chi} \end{equation} where $\Gamma$ is the Gamma function,
$\beta_0=<\beta>$ and the number of degrees of freedom $\nu$ can be found from:
\begin{equation} \nu=\frac{2<\beta>^2}{<\beta^2>-<\beta>^2}. \label{nu}
\end{equation}

The Tsallis parameter $q$ satisfies \cite{tsallis_beck1} 
\begin{equation} q\equiv
1 + \frac{2\alpha}{\alpha \nu +1}. \label{q} \end{equation}

To justify our assumption that the 'local' volatility of the normalized log
returns $Z(t,\Delta t)$ is of the form of $\chi^2$-distribution, we checked the
distribution of the normalized log returns of the daily closing price 
of S\&P500.
We have calculated the standard deviation of the normalized log returns within
various non-overlapping windows of size $m$, ranging from 25 to 1000 ~days

\begin{equation} \beta(k)=\sqrt{{1 \over m}\sum_{i=km+1}^{(k+1)m}Z^2(i) -
\left({1 \over m}\sum_{i=km+1}^{(k+1)m}Z(i)\right)^2} \label{volatility}
\end{equation}

In doing so we have a various number of ${\cal M}$ non-overlapping windows for
various time lags $\Delta t$, and have searched for the most efficient size of
the window in order not to loose data points and therefore information. The
resulting empirically obtained distributions of the 'local' volatility
(Eq.(\ref{volatility})) of normalized log returns for the different different
time lags of interest are plotted in Fig. 7 for an intermediary case 
$m=32$. The
values of the degree $\nu$ of the $\chi^2$-distribution are then obtained using
Eq.~(\ref{nu}). The spread $[\beta_{min},\beta_{max}]$ of the local volatility
$\beta$ decreases with increasing the time lag as it is expected from a
$\chi^2$-distribution function due to the exponential function in Eq. 
(\ref{chi})
for large values of the degree of freedom $\nu$. The value of $\nu$ much varies
as a function of $m$ and the time lags considered. The fits are 
always excellent.
However the  $\beta_0$ and  $\nu$ values are quite dependent on the parameters
used in the numerical analysis. Based on these results, e.g. Fig. 7, it can be
accepted that the (turbulent market) model $\beta$-distributions can be
sufficiently well fitted for our purpose with a $\chi^2$-distribution, thereby
justifying the initial assumption.\footnote{Sattin formula \cite{Sattin2} might
also be tested in future work.}

In order to investigate the impact of the $\alpha$ parameter on the 
tail behavior
of the Tsallis type distribution function we tested Eq. 
(\ref{tsallis}) for fixed
$q$ in two cases : for a time lag $\Delta t=1$~day and $q=1.39$ (Fig. 
8a) and for
a time lag $\Delta t=40$~days and $q=1.22$  for $\alpha=1,0.9,0.8$ (Fig. 8b).
Next we tested Eq. (\ref{tsallis}) for fixed $\alpha=1$ and varying $q$ : for a
time lag $\Delta t=1$~day and for $q=3/2,7/5,4/3$ (Fig. 8c) and for a time lag
$\Delta t=40$~days  for $q=7/5,4/3,5/4$ (Fig. 8d). As expected the tails of the
distribution functions approach a Gaussian type when $q$ is approaching 1. For
completeness, the corresponding cases of the distribution of price 
increments are
shown  and briefly discussed in the Appendix.

In so doing the probability distributions of the normalized log returns for the
different values of the time lag $\Delta 
t=1,5,10,15,20,25,30,35,40$~days can be
shown in Fig. 2 together with the lines representing the best fit to 
the Tsallis
type of distribution function. In Table \ref{table3} the statistical parameters
related to the Tsallis type of distribution function are summarized, 
including a
criterion for the goodness of the fit, i.e. the Kolmogorov-Smirnov distance
$d_{KS}$, which is defined as the maximum distance between the cumulative
probability distributions of the data and the fitting lines. Note that the
kurtosis (see Table \ref{table3}) for the Tsallis type of distribution function
\begin{equation} K_r=K_L\frac{(5-3q)}{(7-5q)}, \label{kr} \end{equation} where
$K_L=3$  for a Gaussian process, is positive for all values of $q<7/5$ as
expected, since its positiveness is directly related to the occurrence of
intermittency \cite{neto}. Moreover, the limit $q<7/5$ also implies that the
second moment of the Tsallis type distribution function will always remain
finite, as necessarily due in the type of phenomena hereby studied. 
Furthermore,
if we assume that the Kolmogorov log-normal model of turbulence \cite{k62} is
applicable and let $\Delta t_L$ be the scale at which the $whole$ partial
distribution function becomes Gaussian, then the kurtosis $K_r$ should scale as

\begin{equation} K_r=K_L\left(\frac{\Delta t}{\Delta t_L}\right)^{-\delta}
\label{krL}. \end{equation}  Therefore

\begin{equation} q=\frac{5 - 7 \left(\Delta t/\Delta 
t_L\right)^{-\delta}} {3 - 5
\left(\Delta t/\Delta t_L\right)^{-\delta}}. \label{qkr} \end{equation}

In order to obtain an estimate for $\Delta t_L$, we observe that the turbulence
model, Eq. (\ref{tsallis}), fits well the normalized log returns for $\Delta
t=120$~days and $q=1.01$ (Fig. 9a). The $\alpha$-parameter ($\alpha=0.74$) in
this case plays an important role in controlling the tails such that 
the Tsallis
type distribution function for negative values of $Z$ fits the data whose
probability distribution function still deviates from Gaussian. In 
fact, further
increasing  the time lag to the value $\Delta t=200$~days leads to a complete
coincidence between the distribution functions in the Tsallis and 
Gaussian forms
for the presently investigated data (Fig. 9b). Corresponding 
parameter values are
also listed in Table \ref{table3}. This short observation 
convincingly indicates
where the transition occurs between the small time scale  model of 
nonextensive,
intermittent process and the large scale  Gaussian extensive homogeneous
fluctuation picture \cite{neto,tsallis} and refine the estimate of the Gaussian
range in Figs.1-2.

In Fig. 10 the Tsallis parameter $q$ is shown as a function of the 
rescaled time
lags $\Delta t/\Delta t_L$, where $\Delta t_L$ is the integral scale, the scale
at which the $whole$ probability distribution function converges to 
Gaussian. The
crosses) represent the $q$ values for which the best fit to the S\&P500 data
(Fig. 2) is obtained with Eq. (\ref{tsallis}). With this the value of the
integral scale $\Delta t_L$, we find the value of the exponent 
$\delta= 0.39$ as
the one for which the Eq. (\ref{qkr}) fits best the $q$-values. The exponent
value $\delta= 0.39$ also allows to fit well the power law dependence (Eqs.
(\ref{kr}) and (\ref{krL})) of the rescaled kurtosis $K_r/K_L$ as shown in the
inset of Fig. 10.

Note that in the framework of the Kolmogorov log-normal model
\cite{k62,tsallis_beck2}, $\delta= 4\kappa/9$, where $\kappa$ is called the
intermittency exponent. Therefore, we find $\kappa=0.88$ for the intermittency
exponent of normalized log returns of the S\&P500 daily closing price 
in the time
interval of interest. This value of $\kappa$ is higher than the value of the
intermittency exponent $\kappa=0.25$ for turbulence recently obtained from
experimental atmospheric data \cite{sree_update}. Early estimates have varied
from 0.18 to 0.85  using different experimental techniques
\cite{sree_temp,sree_mf,wyngaard}. Large values of the intermittency exponent,
ranging from 0.2 to 0.8, have been reported in studies of multiparticle
production \cite{janik}. It was found that the range of intermittency exponent
values  depend on the number of cascades; the smaller the number of 
stages of the
multiplicative cascade the smaller $\kappa$, and conversely [Fig. 2b in
\cite{janik}]. In analogy with such findings, a value of $\kappa=0.88$ can be
considered to be related to a high number of cascades in a multiplicative
process, leading to the observed partial distribution functions of 
the normalized
log returns of  the S\&P500 index.

One can explore the Tsallis type of the probability distribution function
Eq.(\ref{tsallis}) in two limits. For small values of normalized log 
returns $Z$
the probability distribution function converges to the form

\begin{equation} p_{\Delta t}(Z) \approx \frac{1}{Z_q}\exp\left\{ -\frac{C
\beta_02\alpha}{2\alpha-(q-1)}|Z|^{2\alpha}\right\} \label{ts_gauss}
\end{equation} Therefore the Tsallis type distribution function converges to a
Gaussian, i.e. $\alpha \longrightarrow 1$, for small values of the 
normalized log
returns, for any $\Delta t$ investigated hereby (see Figs. 1-2). It is also of
interest to check the probability of return to the origin $p_{\Delta t}(Z=0)$
(Table \ref{table3}). There is a slight difference between the values of the
probability of ''return to the origin'' for the data and the one obtained from
Eq.(\ref{tsallis}) $p_{\Delta t}(Z=0)=1/Z_q$. This difference is 
decreasing with
increasing $\Delta t$ and completely disappears in the Gaussian limit
$q\longrightarrow 1$, $\alpha \longrightarrow 1$.

In the limit of large values of normalized log returns $Z$, the Tsallis type
distribution converges to a power law

\begin{equation} p_{\Delta t}(Z) \approx 
\frac{1}{Z_q}\left\{\frac{(q-1)C \beta_0
2\alpha}{2\alpha-(q-1)}|Z|^{2\alpha}\right\}^{-\frac{1}{q-1}.} \label{ts_power}
\end{equation}

Studying the Tsallis type of distribution function one can obtain from
Eq.(\ref{tsallis}) an expression for the width of the Tsallis type of 
probability
distribution function, $2\sigma_w^2=(2\alpha-(q-1)) / (2\alpha C\beta_0(q-1))$.
In the limit of $\alpha \longrightarrow 1$ the width of the Tsallis type
distribution $2\sigma_w^2=(3-q)/2C\beta_0(q-1)$, i.e. $\sim 
2/(C\beta_0)$. It is
obvious that for large time lags $2\sigma^2_w$ tends to diverge \cite{michael},
like $\simeq (\Delta t)^{2/(3-q)}$; this can be easily verified  on a log-log
plot (not shown).

In limit of $q\longrightarrow 1$ the Tsallis type distribution 
function converges
to Gaussian (as seen in Fig. 9b). The values of the parameters $q$, $\alpha$,
$C\beta_0$, that best fit the data using Eq.(\ref{tsallis}), and $2\sigma_w^2$
are plotted as a function of  the time lag in Fig. 11.

\section{Fokker-Planck approach}

On the other hand, the evolution of a time dependent probability distribution
function is usually  described within the Fokker-Planck approach. This method
provides some further information on the correlations present in the 
time series
and it begins with the joint PDF's, that depend on $\cal{N}$ variables, i.e.
$p^{\cal{N}} (Z_1,\Delta t_1;...;Z_{\cal{N}},\Delta t_{\cal{N}})$. We 
started to
address this issue by determining the joint PDF for ${\cal{N}}=2$, i.e.
$p(Z_2,\Delta t_2; \Delta x_1, \Delta t_1)$. The symmetrically tilted character
of the joint PDF contour levels (Fig. 12) around an inertia axis with slope 1/2
points out to some statistical dependence, i.e. a correlation, between the
normalized log returns $Z(t,\Delta t)$ of the daily closing price signal of
S\&P500. A lack of correlations would put the inertia axis on the main diagonal
(Fig. 12).

The conditional probability function is

\begin{equation} p ( Z_{i+1},\Delta t_{i+1}|Z_{i},\Delta t_{i}) =
\frac{p(Z_{i+1},\Delta t_{i+1};Z_{i},\Delta t_{i})}{p(Z_{i}, \Delta t_{i})}
\end{equation} for $i = 1,...,{\cal{N}}-1$. For any $\Delta t_{2}$ $<$ $\Delta
t_{i}$ $<$ $\Delta t_{1}$, the Chapman-Kolmogorov equation is a necessary
condition of a Markov process, one without memory but governed by probabilistic
conditions

\begin{equation} p(Z_{2},\Delta t_{2}|Z_{1},\Delta t_{1})= \int
d(Z_{i})p(Z_{2},\Delta t_{2}|Z_{i},\Delta t_{i})p(\Delta x_{i},\Delta
t_{i}|Z_{1},\Delta t_{1}). \end{equation}

The Chapman-Kolmogorov equation when formulated in $differential$ form yields a
master equation, which can take the form of a Fokker-P1anck equation
\cite{ernst}. Let $\tau=log_2(200/\Delta t)$,

\begin{equation} \frac{d}{d\tau}p(Z,\tau )=\left[-\frac{\partial }{\partial Z}
D^{(1)}(Z,\tau )+\frac{\partial }{\partial Z^{2}} D^{(2)}(Z,\tau 
)\right]p(Z,\tau
) \label{efp} \end{equation} in terms of a drift $D^{(1)}$($Z$,$\tau $) and a
diffusion coefficient $D^{(2)}$($Z$,$\tau $) (thus values of $\tau $ 
represent $
\Delta t_{i}$, $i=1,...$).

The coefficient functional dependence can be estimated directly from 
the moments
$M^{(k)}$ (known as Kramers-Moyal coefficients) of the conditional probability
distributions:

\begin{equation} M^{(k)}=\frac{1}{\Delta \tau }\int dZ^{^{\prime }} 
(Z^{^{\prime
}}-Z)^{k}p(Z^{^{\prime }},\tau +\Delta \tau |Z,\tau ) \end{equation}

\begin{equation} D^{(k)}(Z,\tau )=\frac{1}{k!}\mbox{lim} M^{(k)} \end{equation}
for $\Delta \tau \rightarrow 0$. According to Fig. 13a the drift coefficient
$D^{(1)}\approx0$ and the diffusion coefficients $D^{(2)}$ is well represented
(Fig. 13b) by a parabola

\begin{equation} D^{(2)}(Z)=0.26 Z^2 - 0.005 Z + 0.02 \label{d2} \end{equation}
in the interval $Z\in[-0.175,0.225]$, - noticing that it is smaller 
than the one
presented in Fig.2.

It may be worthwhile to recall that the observed quadratic dependence of the
diffusion term $D^{(2)}$ is essential for the logarithmic scaling of the
intermittency parameter in studies on turbulence.

Finally, the Fokker-Planck equation for the distribution function is 
known to be
equivalent to a Langevin equation for the variable, i.e. $Z$ here, (within the
Ito interpretation \cite{risken,ernst,reich,hanggi,gardiner})

\begin{equation} \frac {d}{d\tau} Z(\tau) = D^{(1)}(Z(\tau),\tau) + \eta(\tau)
\sqrt {{D^{(2)} (Z(\tau),\tau)}}, \label{lee} \end{equation} \noindent where
$\eta(\tau)$ is a fluctuating $\delta$-correlated force with Gaussian 
statistics,
i.e. $<$ $\eta(\tau)$ $\eta(\tau')$$>$ = 2$ \delta (\tau -\tau')$.

Thus the Fokker-Planck approach provides the evolution process of 
PDF's {\it from
small time lags to larger ones}. The fact that the drift coefficient is
approximately equal to zero, therefore indicating that there is no correlation
between the probability distribution functions for different time lags, is well
related to the Gaussian character of the distribution function for such small
values of the normalized log returns $Z\in[-3,3]$. $D^{(1)}\approx 0$ further
implies that there is almost no "restoring force", i.e. 
$\gamma\approx 0$ in Eq.
(\ref{force}), while the quadratic dependence of $D^{(2)}$ in $Z$ is obviously
like an autocorrelation function for a diffusion process.

\section{Conclusion}

In summary, we have presented a method that provides the evolution process of
probability distribution functions (over twenty years) of one financial index,
i.e. the S\&P500. We have studied the evolution process of the tails that are
outside the central (Gaussian) regime at small returns, thereby 
facilitating the
understanding of the evolution of these distribution functions in a 
Fokker-Planck
framework. Beck turbulence model can be well applied to describe the volatility
(of normalized log-returns) distributions assuming  a $\chi^2$-distribution for
the ''local'' volatility. An  open question in nonextensive thermostatistics
studies is often raised about the meaning, value and behavior of the non
extensive exponent, or Tsallis parameter $q$. The intermittency exponent  is
found to be related to the scaling exponent of the PDF moments in the framework
of  Kolmogorov log-normal model, thereby giving weight to the model and the
statistical approach. The intermittency exponent large value  points to a large
number of cascades in the turbulent process. Its range has been found to extend
up to $ca.$ 200 days. One may still wonder on the $q$-value itself.  In other
works, this value is related, e.g. to the upper and lower bounds of the
multifractal dimension \cite{tsallis_wilk}, in other words to the bounds of the
$\alpha$ values in multifractal studies \cite{mf}. It may also be 
related to the
value of the fractional derivative, say in a non-linear Fokker-Planck equation
approach \cite{NLFPEfractderiv}. This should be some interesting work 
to pursue,
- again with some warning concerning the possible error bars on the 
generalized
fractal dimension in multifractal studies \cite{luxwarning}.

We have also presented the turbulence-like dynamics  through the Fokker-Planck
and the Langevin equations. We have (unexpectedly) found that, in the treated
case, there is almost no "restoring force", i.e. ($\gamma\approx 0$ in the
Langevin equation).  A comparison is made between normalized 
log-returns and mere
price increments. We have examined the corresponding cases of the 
distribution of
price increments with other possible definitions. It was found that the
definition (through a normalized log return rather than a mere price 
difference)
is very relevant for obtaining nice fits. This has been also observed in a work
by   Karth and  Peinke \cite{AksoeFriedrich} on related matter. This warning
might also shine some light on the possible origin of the controversy
\cite{3,gopi} concerning the relationship (or not) between the fat 
tails  caused
by  some dynamical hierarchical cascade process of volatility correlations.

These points not withstanding,  we  have related a financial market behavior to
Tsallis non extensive thermodynamics approach, i.e. more precisely to a
turbulence-like process, - as financial market and indices were often 
claimed to
be seen \cite{stanleybook,ghash,schmitt}. Finally, it seems that we have
thoroughly answered the often raised question "why to look at the tails of a
probability distribution function? and what does that lead to?".

\section{Appendix}

We have also searched for describing the partial distribution function of the
(raw) increments of daily closing price signal of the S\&P500 with the Tsallis
type distribution function. We have applied Eq.(\ref{tsallis}) for $x\equiv
\Delta y(t,\Delta t)=y(t+\Delta t) - y(t)$. We have tested the Tsallis type
distribution function for the increments of $\Delta t=1$~day for fixed $q=1.45$
($C\beta_0=0.23$) and varying $\alpha=0.5,0.6,0.7$ (Fig. 14a). 
Applying the same
set of parameters and to price increments for $\Delta y=40$~days, leads to a
pretty bad fit (Figs. 14b). Decreasing the value of $q$ would not have produce
better results since the Tsallis type distribution function would have been
bounded within smaller range around $\Delta y=0$ values. A test for fixed
$\alpha=0.5$ ($C\beta_0=0.23$) and varying $q=1.45,1,30,1.15$ for  $\Delta
t=1$~day is next shown in Fig. 14c. Again the same set of parameters is applied
to price increments for $\Delta y=40$~days and lead to a pretty bad fit (Figs.
14d). These results may be somewhat expected because the Tsallis type
distribution function represents a mathematical construction that is 
designed for
normalized variables, i.e. a variable changing within a limited range. To take
into account a double peak like structure (e.g. for large time lags, 
see Fig.14)
remains an open question.

\vskip 1cm \newpage \parindent=0pt \newpage \parindent=0pt

{\large \bf Figure Captions}

\vskip 0.5cm{\bf Figure 1} -- Probability distribution function of 
normalized log
returns $Z(t,\Delta t)$ of daily closing price value signal of S\&P500 between
Jan. 01, 1980 and Dec. 31, 1999, for $\Delta t=1$~day (symbols). Normalized log
returns are calculated as $Z(t,\Delta t)=(\tilde y(t) - <\tilde y>_{\Delta
t})/\sigma_{\Delta t}$, where $\tilde y(t)=ln(y(t+\Delta t)/y(t))$ and
$\sigma_{\Delta t}$ is the standard deviation of $\tilde y(t)$ for time lag
$\Delta t$. The dashed line represents a Gaussian distribution. 
Inset: Power law
fit (solid line) of the negative (-3.1) and positive (-3.1) slope of the
distribution outside the Gaussian regime

\vskip 0.5cm{\bf Figure 2} -- Probability distribution function (PDF) 
$p_{\Delta
t}(Z)$ of normalized log returns of the daily closing price value signal of
S\&P500 (symbols) and the Tsallis type distribution function (lines) for
different values of $\Delta t=1,5,10,15,20,25,30,35,40$~days. The PDF (symbols
and curves) for each $\Delta t$ are displaced by 10 with respect to 
the previous
one; the curve for $\Delta t=1$~day is unmoved. The large circles mark the ends
of the interval in which the distribution is $like$  a gaussian 
distribution. The
values of the slopes of the positive and negative tails of the distributions
outside the gaussian range $Z\in [-3,3]$ are listed in Table \ref{table1}. The
values of the parameters for the Tsallis type distribution function for each
$\Delta t$ are summarized in Table \ref{table3}

\vskip 0.5cm{\bf Figure 3} -- DFA function $F(n)$ plotted as a function of the
the box size $n$ of the integrated daily closing price value signal of S\&P500
between Jan. 01, 1980 and Dec. 31, 1999. Brownian-like fluctuations with
$1+H_{DFA}=1.52\pm0.01$ are obtained for all possible time scales

\vskip 0.5cm{\bf Figure 4} -- DFA function $F(n)$ plotted as a function of the
the box size $n$ of the integrated normalized log returns $Z(t,\Delta t)$ of
daily closing price value signal of S\&P500 between Jan. 01, 1980 and Dec. 31,
1999 for different time lags $\Delta 
t=1,5,10,15,20,25,30,35,40$~days. Values of
the scaling exponents $H_{DFA}$ for the various DFA functions are summarized in
Table \ref{table2}. Inset: Power law functional dependence of the value of the
cross over box-size $n_x$ as a function of time lag $\Delta t$

\vskip 0.5cm{\bf Figure 5} -- Power spectrum $S(f)$ of the daily closing price
value signal of S\&P500 between Jan. 01, 1980 and Dec. 31, 1999. A 
scale break at
around $f=1/250$~day$^{-1}$ separates two scaling regions. Inset: 
Scaling of the
power spectrum of the daily closing price signal of S\&P500 as a white noise
signal with $\mu\approx 0$

\vskip 0.5cm{\bf Figure 6} -- Power spectrum $S(f)$ of the normalized 
log returns
$Z(t,\Delta t)$ of daily closing price value signal of S\&P500 between Jan. 01,
1980 and Dec. 31, 1999 for different time lags $\Delta 
t=1,5,20,35,40$~days. Each
curve is displaced by $10^{-5}$ with respect to the previous one; the power
spectrum of the normalized log returns for $\Delta t=1$~day is not 
displaced. The
dashed line from $f=1/70$~days$^{-1}$ to $f=1/2$~days$^{-1}$ has a slope
$\mu=1.86$, corresponding to the $H_{DFA}$ exponent. The horizontal dashed line
from $f=1/10^{-4}$~days$^{-1}$ to $f=1/128$~days$^{-1}$ corresponds to what
should be expected for white noise and is in agreement with the  scaling of the
DFA function for the same data in Fig. 4

\vskip 0.5cm{\bf Figure 7} -- Probability density $f_{\Delta t}(\beta)$ of the
local volatility $\beta$ (Eq.(\ref{volatility})) in terms of standard deviation
of the normalized log returns $Z(t,\Delta t)$ of S\&P500 in non-overlapping
windows with size $m$=32~days for different time lags (symbols) (a-i) $\Delta
t=$1, 5, 10, 15, 20, 25, 30, 35, 40~days. Lines: $\chi^2$-distribution as given
by Eq. (\ref{chi}).

\vskip 0.5cm{\bf Figure 8} -- Probability distribution functions of the
normalized log returns of daily closing price signal of S\&P500 
(symbols) for (a)
$\Delta t=1$~day and fixed $q=1.39$. The Tsallis type distribution functions
(Eq.(\ref{tsallis})) obtained for various values of the parameter
$\alpha=1.0,0.9,0.8$, dashed, solid, dash-dotted line, respectively; 
(b) same as
(a) but for $\Delta t=40$~days and $q=1.22$; (c) for $\Delta t=1$~day and fixed
$\alpha=1.0$ for various values of $q=3/2, 7/5, 4/3$, dashed, solid, 
dash-dotted
line, respectively; (d) same as (c) but for $\Delta t=40$~days and $q=7/5, 4/3,
5/4$

\vskip 0.5cm{\bf Figure 9} -- Partial distribution function of normalized log
return of daily closing price of S\&P500 for a large time lag, i.e. (a) $\Delta
t=120$~days and (b) $\Delta t=200$~days. The solid line marks the 
best fit with a
Tsallis type distribution function, Eq. (\ref{tsallis}), while the Gaussian
distribution function is drawn with a dashed line

\vskip 0.5cm{\bf Figure 10} -- The functional dependence of the Tsallis $q$
parameter on the rescaled time lag $\Delta t/\Delta t_L$ for $\Delta
t_L=200$~days and $\delta= 0.39$ (see Eq. (\ref{qkr})) (line);  the symbols
represent the values of the $q$ parameter listed in Table \ref{table3} and used
to plot the fitting lines in Figs. 2 and 9. Inset : Scaling properties of the
rescaled kurtosis $K_r/K_L$, where $K_L=3$ is the kurtosis for a Gaussian
process, as a function of the rescaled time lag $\Delta t/\Delta t_L$ 
satisfying
Eq. (\ref{kr}) (open symbols) and Eq. (\ref{krL}) (full symbols)

\vskip 0.5cm{\bf Figure 11} -- Characteristic parameters of Tsallis type
distribution function as defined in \cite{kozuki} : Tsallis $q$-parameter
(crosses), $\alpha$ (squares), constant $C\beta_0$ used in the fit (open
circles), the width of the Tsallis type distribution 
$2\sigma_w^2=(2\alpha-(q-1))
/ (2\alpha C\beta_0(q-1))$ from Eq.(\ref{tsallis}) (triangles) (rescaled by a
factor of 1/6), asymptotic behavior of $2\sigma_w^2\approx 2/(C\beta_0)$ for
$\alpha\longrightarrow 1$ (full circles) (rescaled by a factor of 1/6)

\vskip 0.5cm{\bf Figure 12} -- Typical contour plots of the joint probability
density function $p(Z_2,\Delta t_2; Z_1,\Delta t_1)$ of daily closing price of
S\&P500 for the period of interest Jan. 01, 1980 and Dec. 31, 1999. 
Dashed lines
have a slope +1 and emphasize the correlations between probability density
functions for (a) $\Delta t_2=1$~day and $\Delta t_1=5$~days and (b) $\Delta
t_2=5$~days and $\Delta t_1=10$~days. Contour levels correspond to
$log_{10}p(Z_2,\Delta t_2; Z_1,\Delta t_1) =-1.0,-1.5,-2.0,-2.5,-3.0$ 
from center
to border

\vskip 0.5cm{\bf Figure 13} -- Kramers-Moyal drift and diffusion 
coefficients (a)
$D^{(1)}$ and (b) $D^{(2)}$ as a function of normalized log returns 
$Z$ for daily
closing price of S\&P500 ; $ D^{(2)} = 0.26 Z^2 - 0.0005 Z + 0.02$

\vskip 0.5cm{\bf Figure 14} -- Probability distribution functions of the daily
closing price increments $\Delta y(t)=y(t+\Delta t)-y(t)$ of S\&P500 (symbols).
The Tsallis type distribution functions (Eq.(\ref{tsallis})) obtained (a) for
$\Delta t=1$~day and fixed $q=1.45$ ($C\beta_0=0.23$) and for various values of
the parameter $\alpha=0.5,0.6,0.7$, dashed, solid, dash-dotted line,
respectively; (b) the same as (a) but for $\Delta t=40$~days; (c) for $\Delta
t=1$~day and fixed $\alpha=0.5$ ($C\beta_0=0.23$) and for various values of
$q=1.45,1.30,1.15$, dashed, solid, dash-dotted line, respectively; (d) the same
as (b) but for $\Delta t=40$~days

\newpage

\begin{table} \caption{Slopes of the positive and negative tails (second and
third column) of the distributions outside the Gaussian range 
$Z(t,\Delta t) \in
[-3,3]$; values of characteristic Tsallis function parameters are given}
\begin{center} \begin{tabular}{rccccc} \hline $\Delta
t$&positive&negative&$2q$&$1/(q-1)$&$\alpha$\\ \hline 1 & 3.1 & 3.1 & 2.78&
2.564& 0.92 \\ 5 & 3.1 & 3.0 & 2.72 & 2.778& 0.90 \\ 10 & 3.0 &2.7 & 2.68 &
2.941& 0.88 \\ 15 & 3.0 & 2.6 & 2.66 & 3.030& 0.86 \\ 20 & 2.9 & 2.5 & 2.64 &
3.125& 0.85  \\ 25 & 2.9 & 2.5 & 2.62 & 3.226&0.83  \\ 30 & 2.8 & 2.3 & 2.58 &
3.448& 0.80 \\ 35 & 2.7 & 2.3 &2.50 & 4.000& 0.78  \\ 40 & 2.5 & 2.2 & 2.44 &
4.545& 0.76 \\ \hline \end{tabular} \end{center} \label{table1} \end{table}

\newpage \begin{table} \caption{Values of the scaling exponent from the DFA
analysis of normalized log returns $Z$ for different values of the time lag
$\Delta t=5,10,15,20,25,30,35,40$~days, and crossover ''box size'' $n_x$ }
\begin{center} \begin{tabular}{rcccc} \hline $\Delta t$& $1+H_{DFA}$ & $n_x$ \\
\hline 5 &      1.27$\pm$0.04 &19\\ 10 & 1.37$\pm$0.02 & 35 \\ 15 & 
1.39$\pm$0.02
& 49 \\ 20 &      1.38$\pm$0.02 & 70\\ 25 & 1.38$\pm$0.02 & 91 \\ 30 &
1.41$\pm$0.01 & 108\\ 35 & 1.43$\pm$0.01 & 117 \\ 40 & 1.43$\pm$0.01 & 128 \\
\hline \end{tabular} \end{center} \label{table2} \end{table}

\begin{table} \caption{ Values of the   parameters characterizing the S\&P500
daily closing price data between Jan. 1, 1980 and Dec. 31, 1999 in  the
nonextensive thermostatisitics approach. For the definition of the
Kolmogorov-Smirnov distance $d_{KS}$ see the text} \begin{center} 
\tabcolsep=4pt
\begin{tabular}{rccccccc} \hline $\Delta t$& $q$ 
&$\alpha$&$C\beta_0$& $p_{\Delta
t}(Z=0)$&$p_{\Delta t}(Z=0)$ & $K_r$&$d_{KS}$ \\ &&&&data&Eq.(4)&&\\ \hline 1
&1.39 & 0.92 & 0.65 & 0.505 & 0.611 & 49.800 & 0.072\\ 5 & 1.36 &0.90 & 0.62 &
0.447 & 0.600 & 13.800 & 0.100\\ 10 & 1.34 & 0.88 &0.60 &  0.462 & 
0.592 & 9.800
& 0.091\\ 15 & 1.33 & 0.86 & 0.58 & 0.472 & 0.582 & 8.657 & 0.085\\ 20 & 1.32 &
0.85 & 0.56 &  0.459 & 0.572 & 7.800 & 0.085\\ 25 & 1.31 & 0.83 & 
0.54 & 0.447 &
0.560 & 7.133 & 0.087\\ 30 & 1.29 & 0.80 & 0.52 &  0.443 &0.549 & 
6.164 & 0.088\\
35 & 1.25 & 0.78 & 0.50 &  0.432 & 0.538 &5.000 & 0.087\\ 40 & 1.22 & 0.76 &
0.48& 0.445 & 0.525 & 4.467 &0.077\\ 120 & 1.01 & 0.74 & 0.39 &  0.431 & 0.467
&3.031 &0.052\\ 200 & 1.01 & 1.00 & 0.26 &   0.398 &  0.406 &  3.031 & 0.040\\
\hline \end{tabular} \end{center} \label{table3} \end{table}


\newpage

\begin{figure}[ht] \begin{center} \leavevmode \epsfysize=10cm
\epsffile{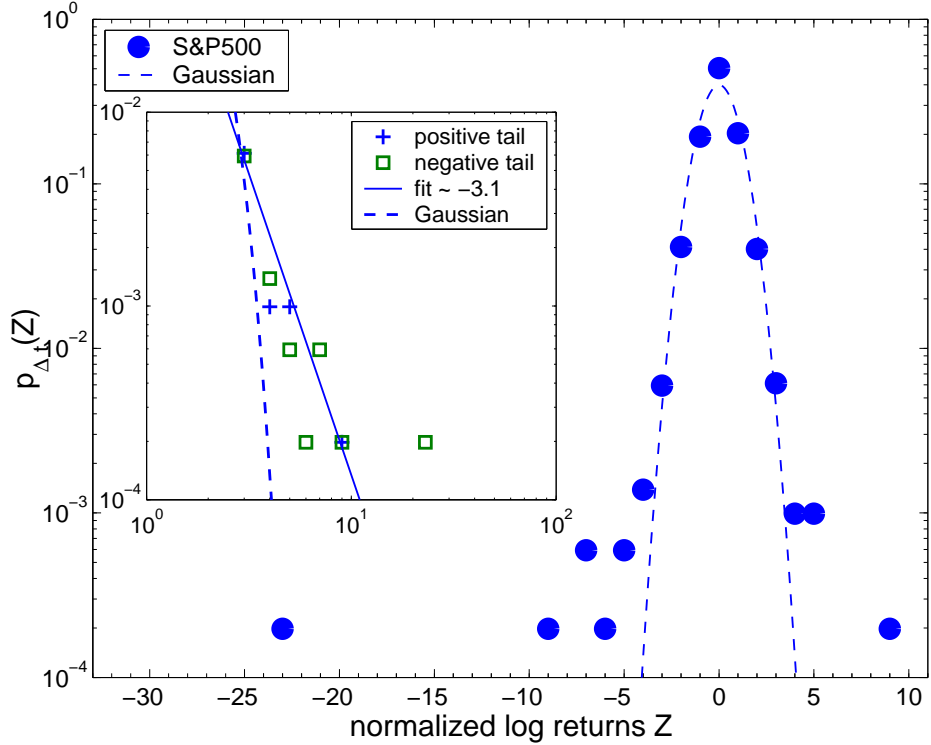} \caption{Probability distribution function of 
normalized log
returns $Z(t,\Delta t)$ of daily closing price value signal of S\&P500 between
Jan. 01, 1980 and Dec. 31, 1999, for $\Delta t=1$~day (symbols). Normalized log
returns are calculated as $Z(t,\Delta t)=(\tilde y(t) - <\tilde y>_{\Delta
t})/\sigma_{\Delta t}$, where $\tilde y(t)=ln(y(t+\Delta t)/y(t))$ and
$\sigma_{\Delta t}$ is the standard deviation of $\tilde y(t)$ for time lag
$\Delta t$. The dashed line represents a Gaussian distribution. 
Inset: Power law
fit (solid line) of the negative (-3.1) and positive (-3.1) slope of the
distribution outside the Gaussian regime} \end{center} \label{fig1}\end{figure}

\newpage \begin{figure}[ht] \begin{center} \leavevmode \epsfysize=9.5cm
\epsffile{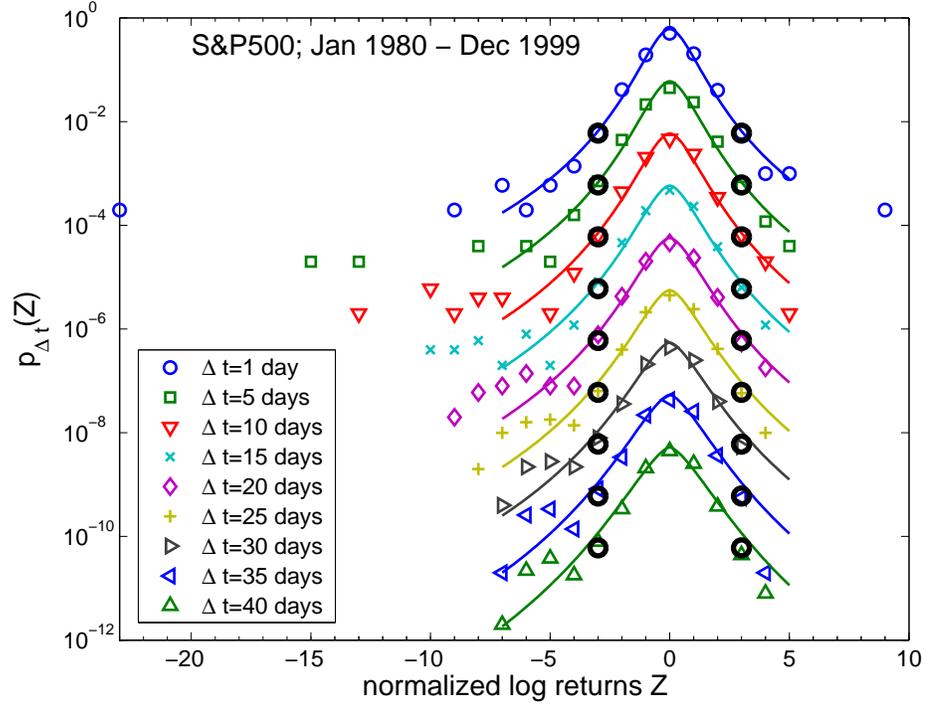} \caption{Probability distribution function (PDF) 
$p_{\Delta
t}(Z)$ of normalized log returns of the daily closing price value signal of
S\&P500 (symbols) and the Tsallis type distribution function (lines) for
different values of $\Delta t=1,5,10,15,20,25,30,35,40$~days. The PDF (symbols
and curves) for each $\Delta t$ are displaced by 10 with respect to 
the previous
one; the curve for $\Delta t=1$~day is unmoved. The large circles mark the ends
of the interval in which the distribution is $like$  a gaussian 
distribution. The
values of the slopes of the positive and negative tails of the distributions
outside the gaussian range $Z\in [-3,3]$ are listed in Table \ref{table1}. The
values of the parameters for the Tsallis type distribution function for each
$\Delta t$ are summarized in Table \ref{table3}} \end{center}\label{fig2}
\end{figure}

\newpage \begin{figure}[ht] \begin{center} \leavevmode \epsfysize=10cm
\epsffile{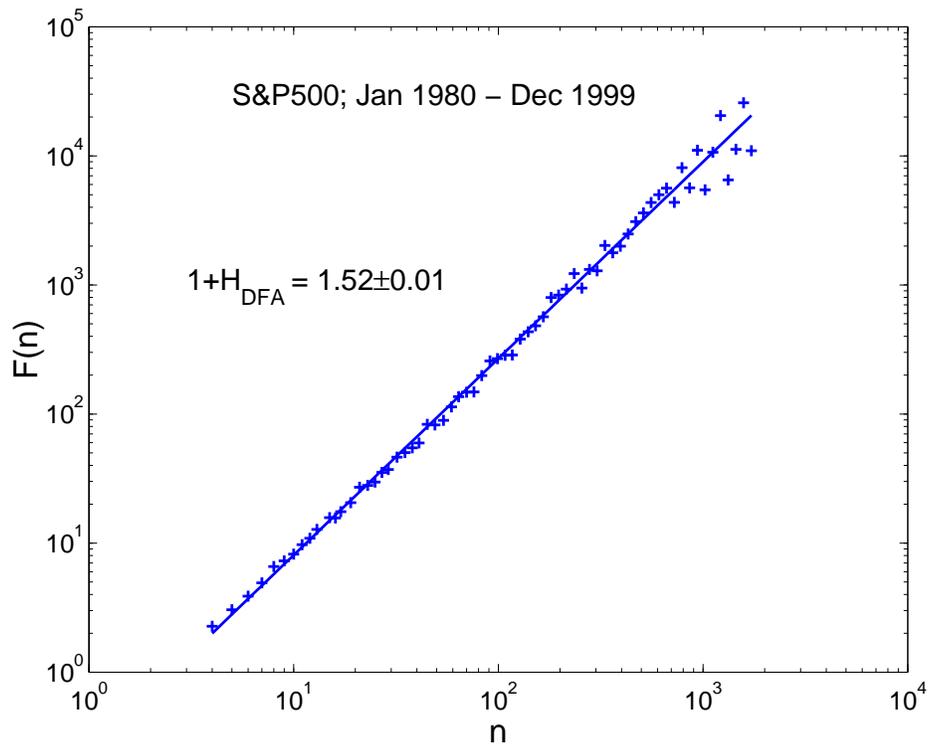} \caption{DFA function $F(n)$ plotted as a function of the
the box size $n$ of the integrated daily closing price value signal of S\&P500
between Jan. 01, 1980 and Dec. 31, 1999. Brownian-like fluctuations with
$1+H_{DFA}=1.52\pm0.01$ are obtained for all possible time scales } 
\end{center}
\label{fig3}\end{figure}

\newpage \begin{figure}[ht] \begin{center} \leavevmode \epsfysize=10cm
\epsffile{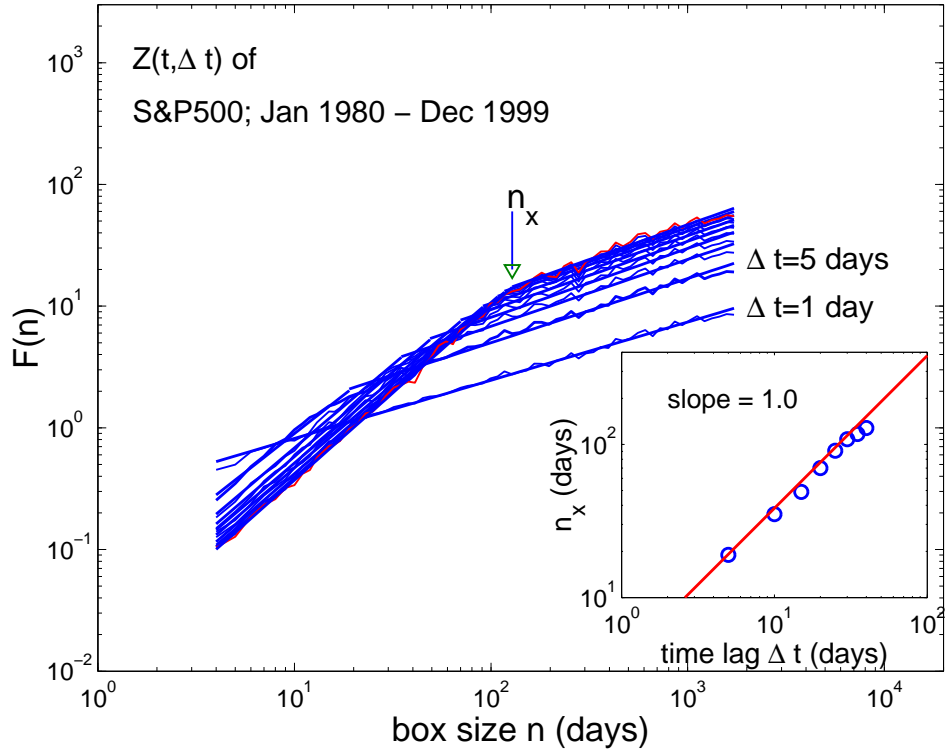} \caption{DFA function $F(n)$ plotted as a function of the
the box size $n$ of the integrated normalized log returns $Z(t,\Delta t)$ of
daily closing price value signal of S\&P500 between Jan. 01, 1980 and Dec. 31,
1999 for different time lags $\Delta 
t=1,5,10,15,20,25,30,35,40$~days. Values of
the scaling exponents $H_{DFA}$ for the various DFA functions are summarized in
Table \ref{table2}. Inset: Power law functional dependence of the value of the
cross over box-size $n_x$ as a function of time lag $\Delta t$} \end{center}
\label{fig4}\end{figure}

\newpage \begin{figure}[ht] \begin{center} \leavevmode \epsfysize=10cm
\epsffile{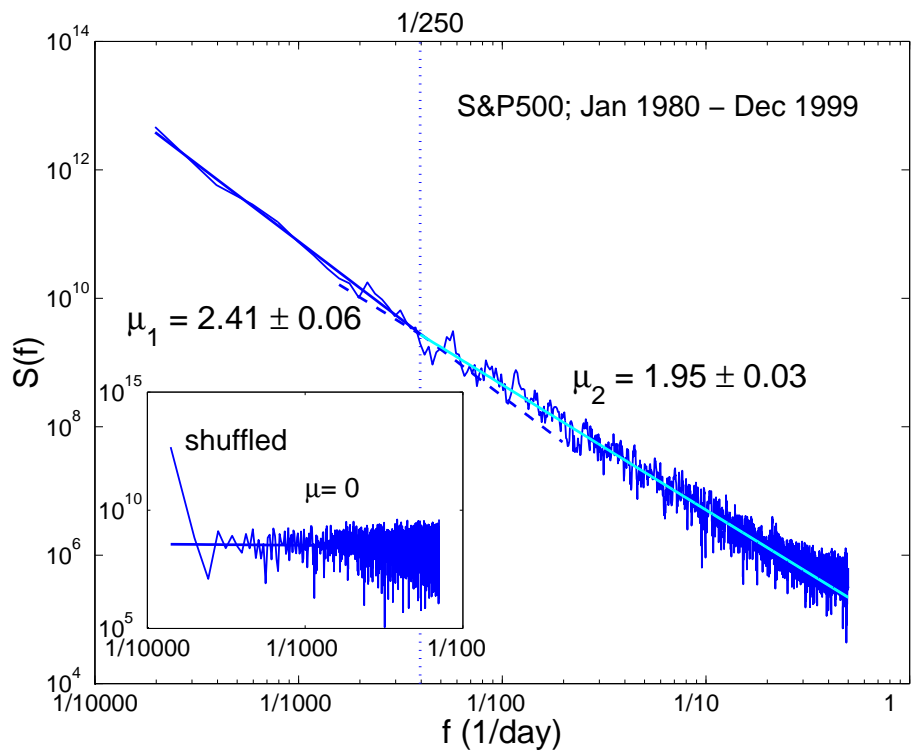} \caption{Power spectrum $S(f)$ of the daily closing price
value signal of S\&P500 between Jan. 01, 1980 and Dec. 31, 1999. A 
scale break at
around $f=1/250$~day$^{-1}$ separates two scaling regions. Inset: 
Scaling of the
power spectrum of the daily closing price signal of S\&P500 as a white noise
signal with $\mu\approx 0$} \end{center} \label{fig5}\end{figure}

\newpage \begin{figure}[ht] \begin{center} \leavevmode \epsfysize=10cm
\epsffile{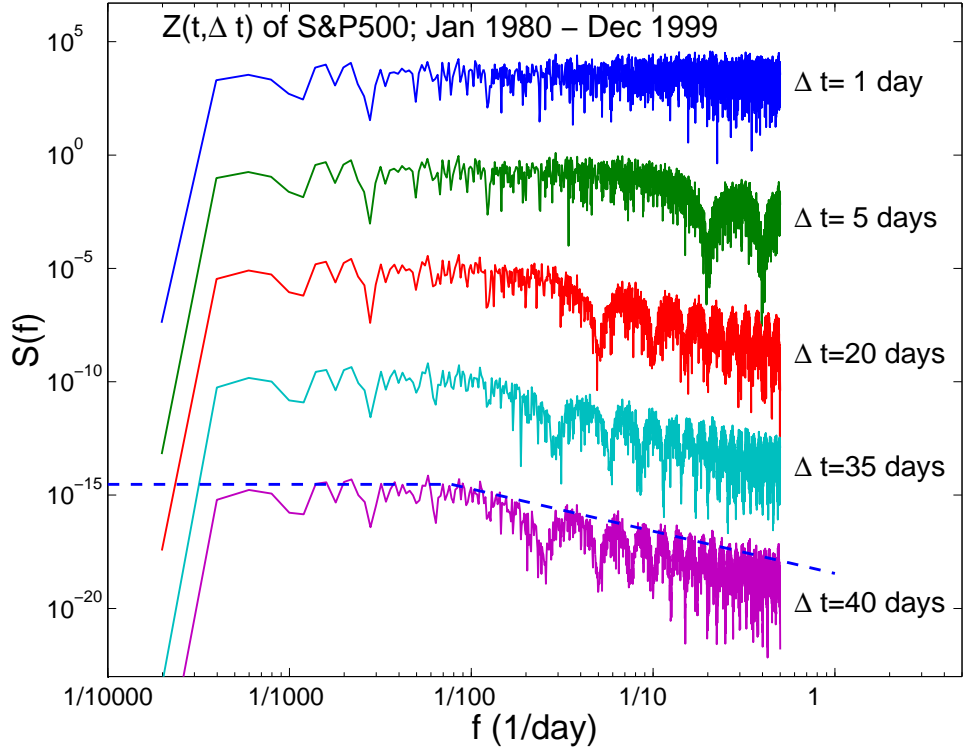} \caption{Power spectrum $S(f)$ of the normalized 
log returns
$Z(t,\Delta t)$ of daily closing price value signal of S\&P500 between Jan. 01,
1980 and Dec. 31, 1999 for different time lags $\Delta 
t=1,5,20,35,40$~days. Each
curve is displaced by $10^{-5}$ with respect to the previous one; the power
spectrum of the normalized log returns for $\Delta t=1$~day is not 
displaced. The
dashed line from $f=1/70$~days$^{-1}$ to $f=1/2$~days$^{-1}$ has a slope
$\mu=1.86$, corresponding to the $H_{DFA}$ exponent. The horizontal dashed line
from $f=1/10^{-4}$~days$^{-1}$ to $f=1/128$~days$^{-1}$ corresponds to what
should be expected for white noise and is in agreement with the  scaling of the
DFA function for the same data in Fig. 4} \end{center}\label{fig6} \end{figure}

\begin{figure}[ht] \begin{center} \leavevmode \epsfysize=3.5cm
\epsffile{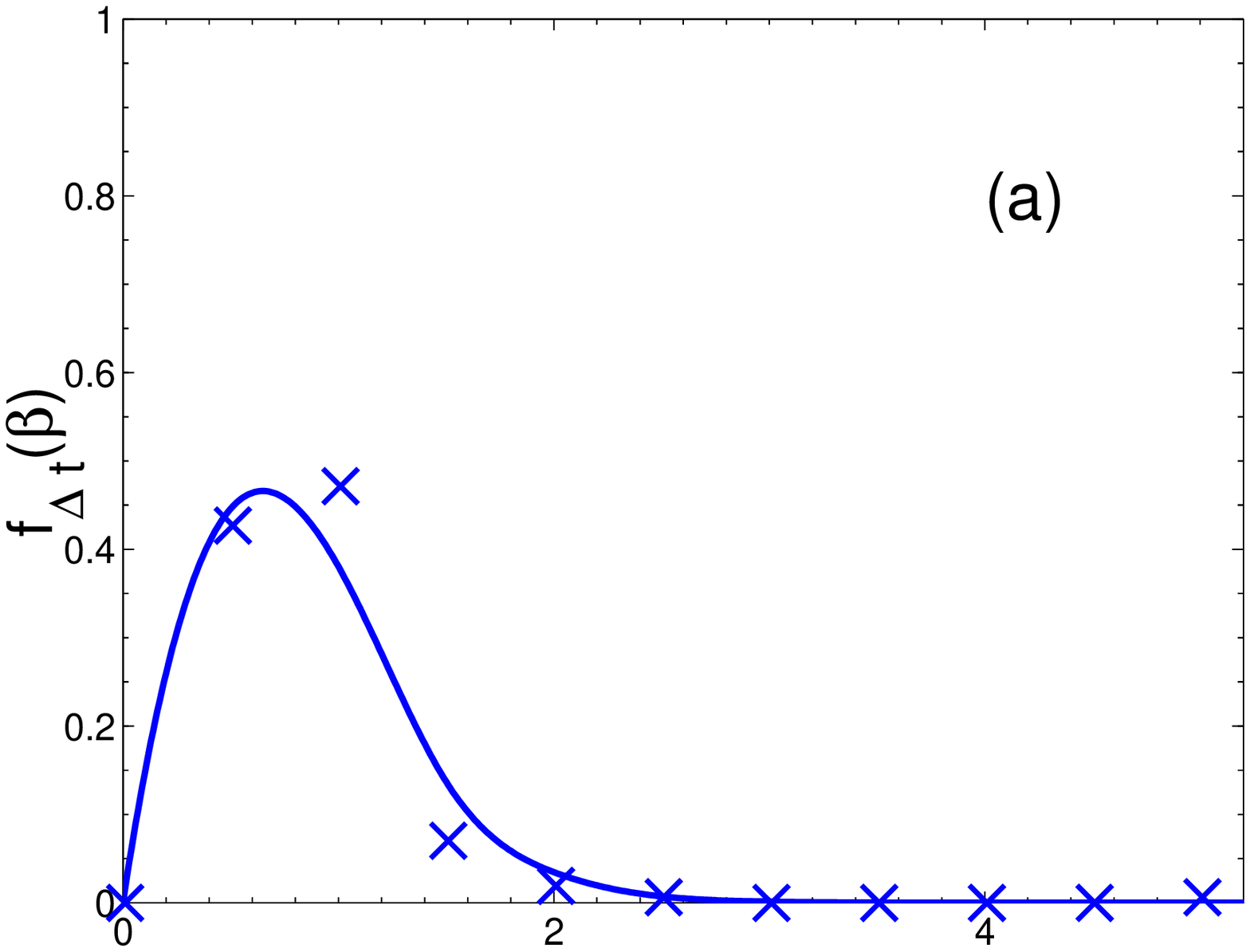} \hfill \leavevmode \epsfysize=3.5cm \epsffile{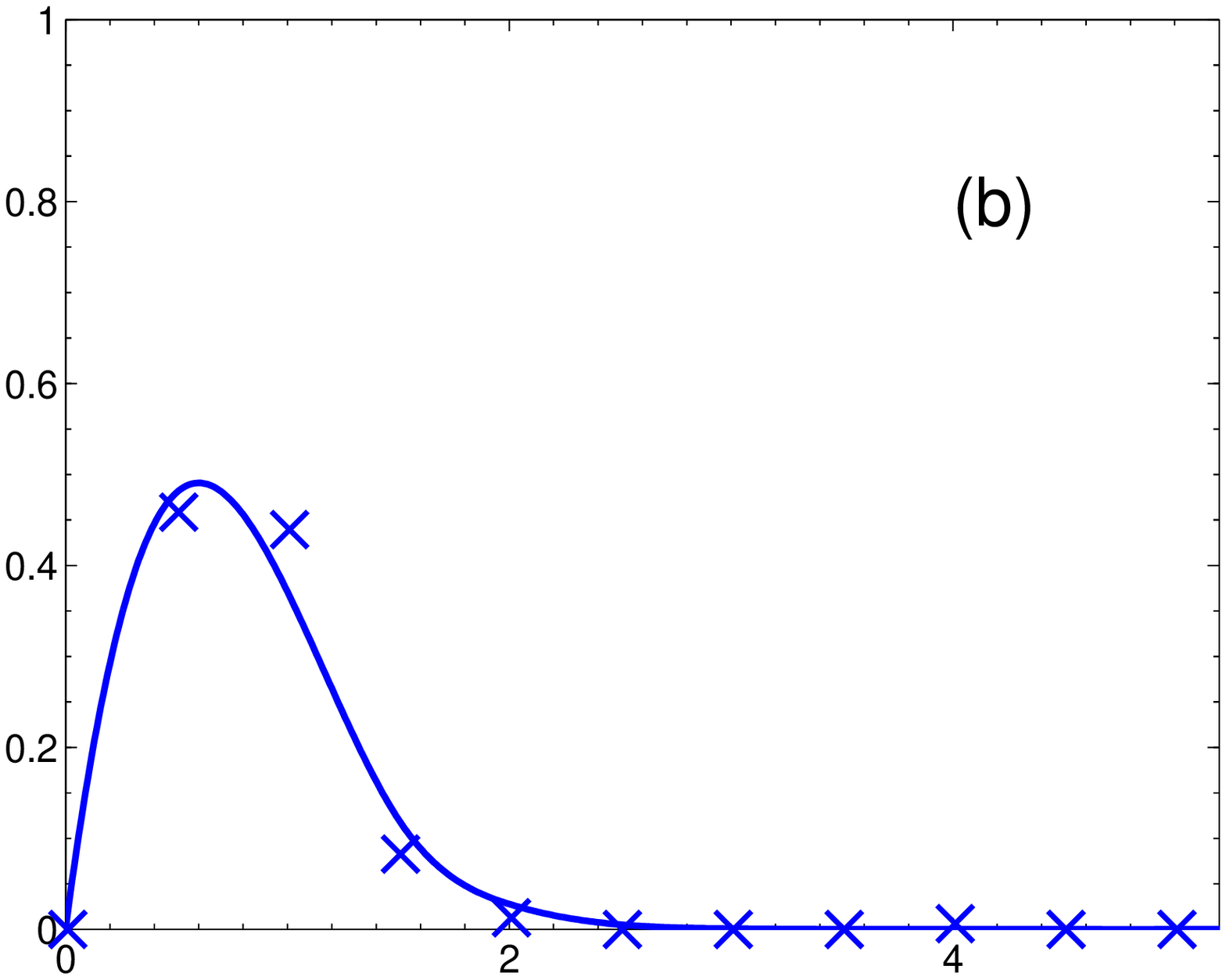}
\hfill \leavevmode \epsfysize=3.5cm \epsffile{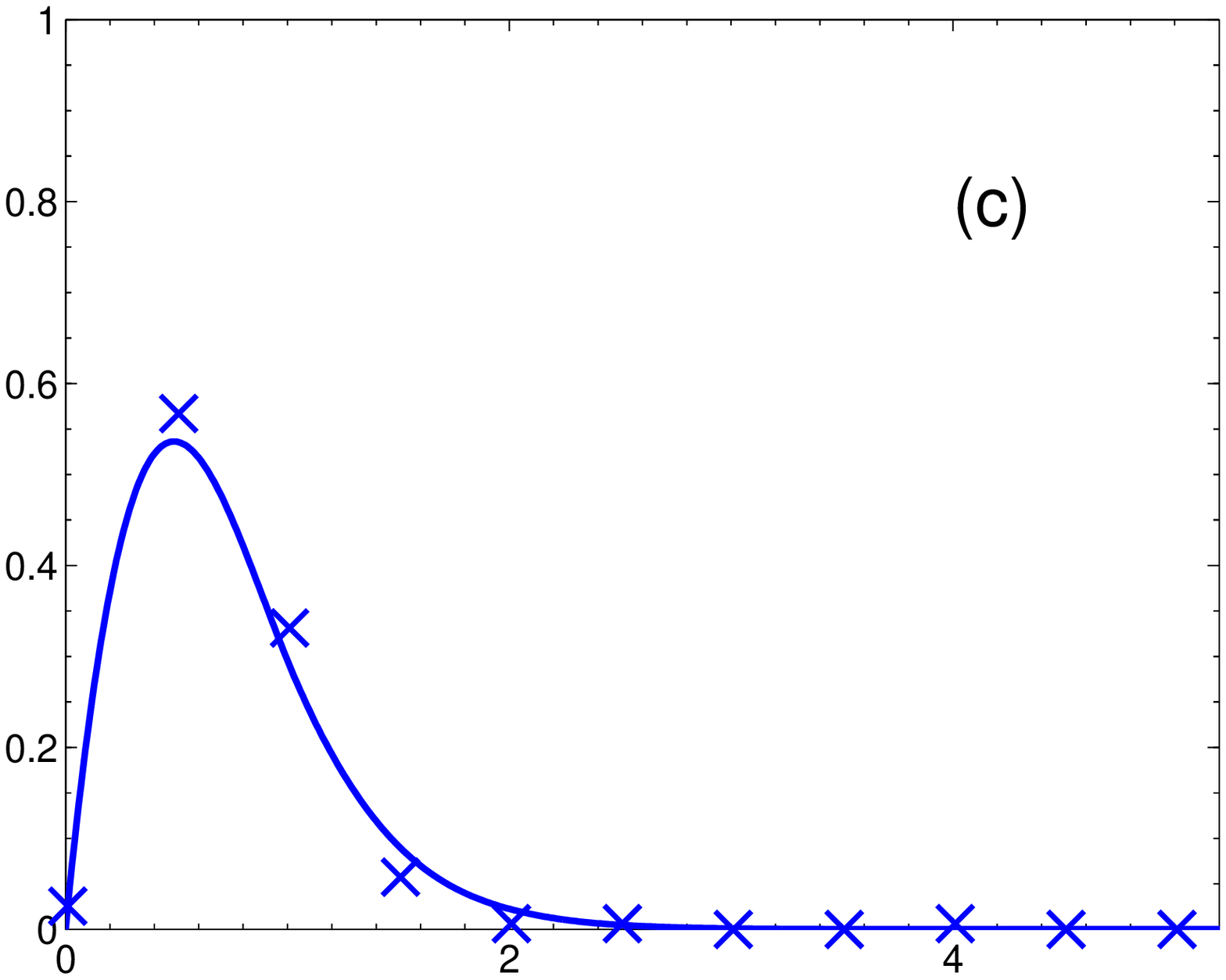} \vfill \leavevmode
\epsfysize=3.5cm \epsffile{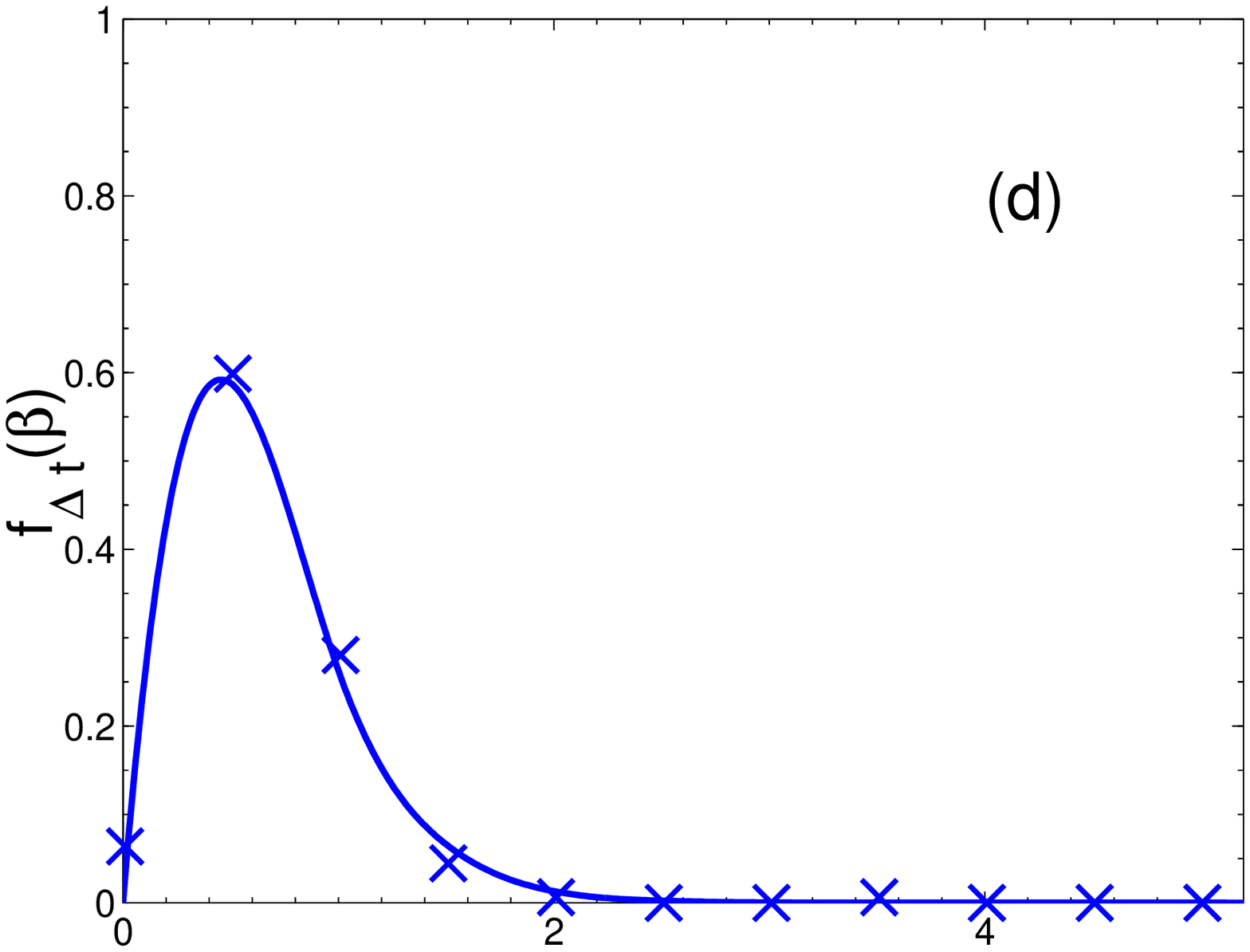} \hfill \leavevmode \epsfysize=3.5cm
\epsffile{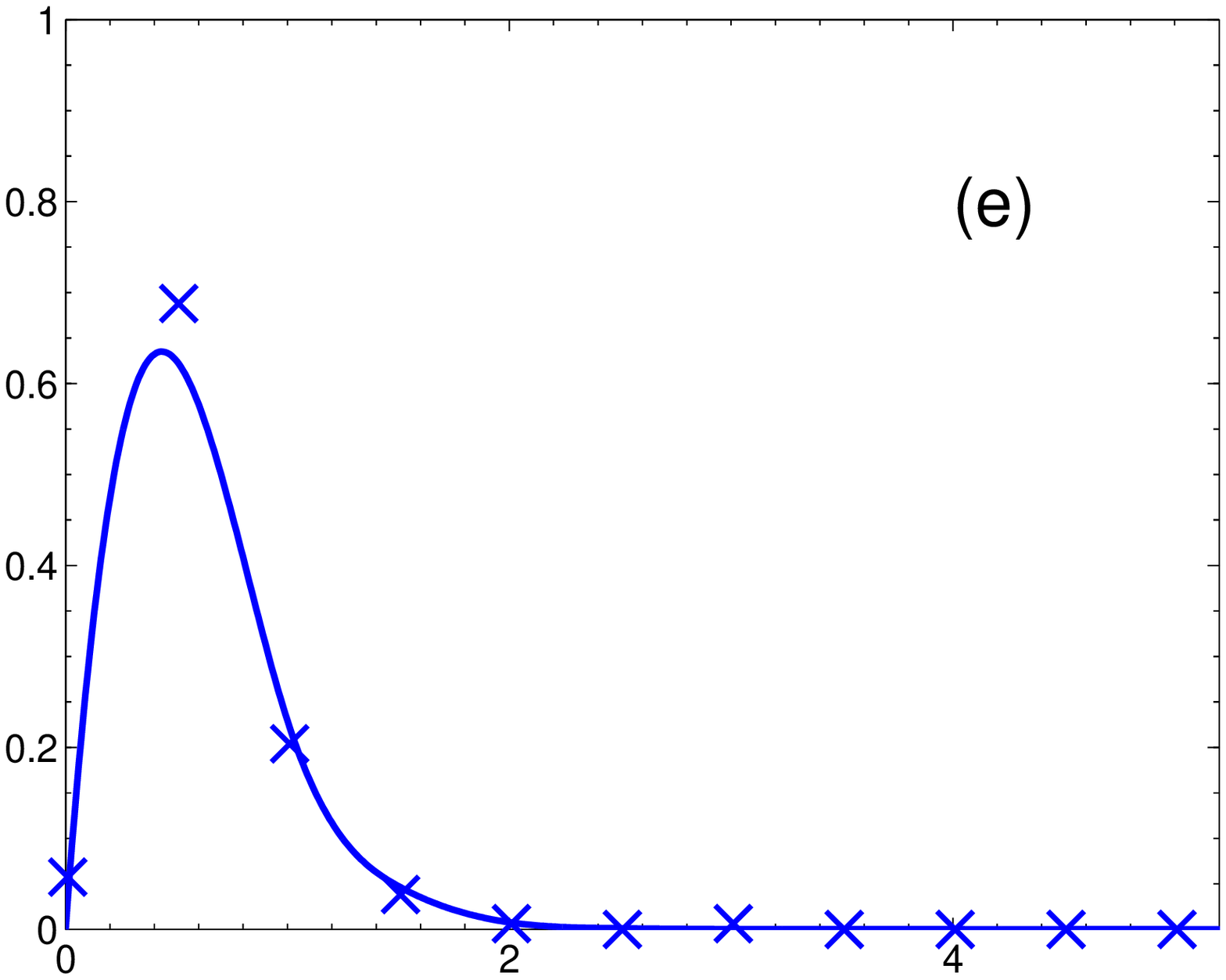} \hfill \leavevmode \epsfysize=3.5cm \epsffile{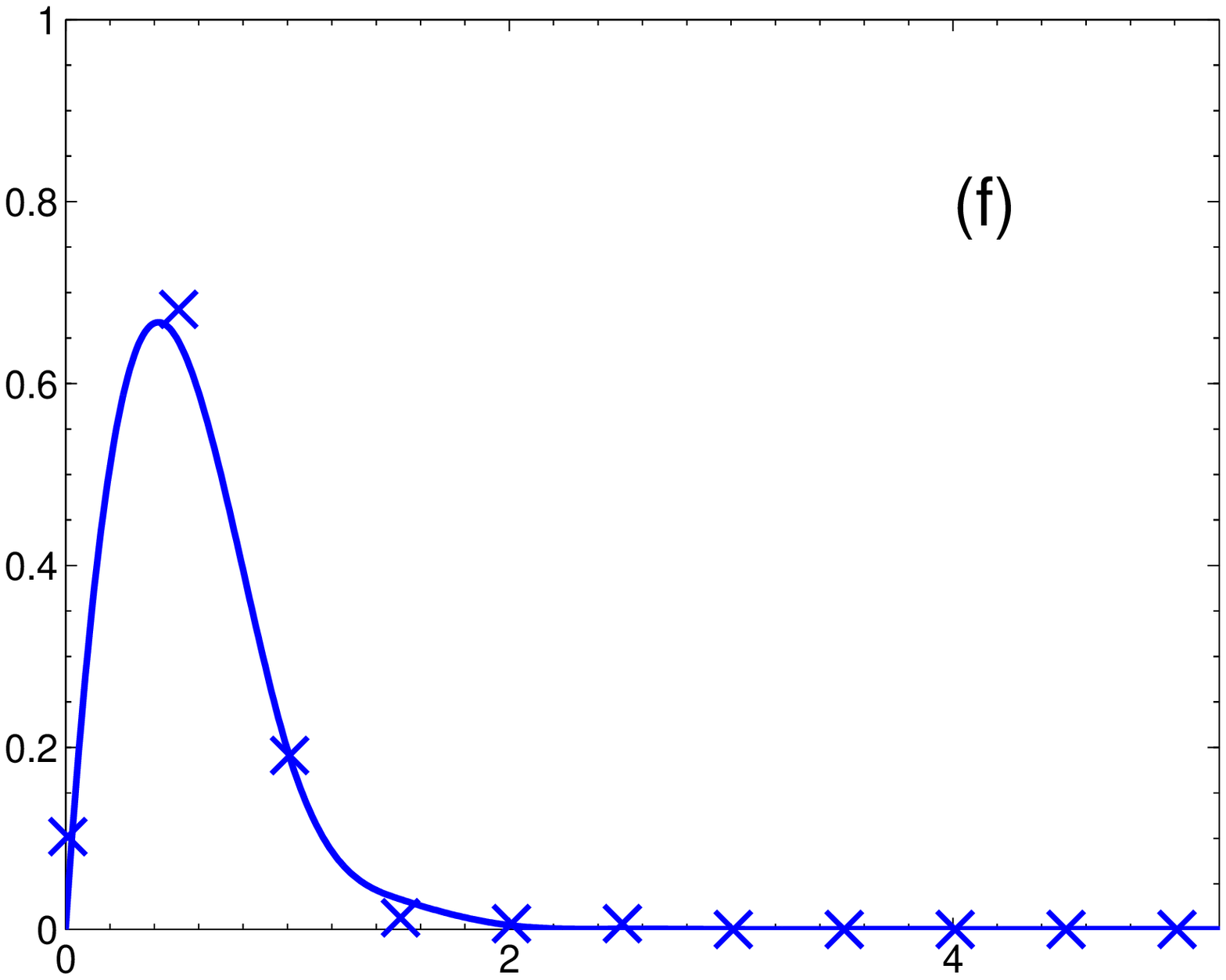}
\vfill \leavevmode \epsfysize=3.7cm \epsffile{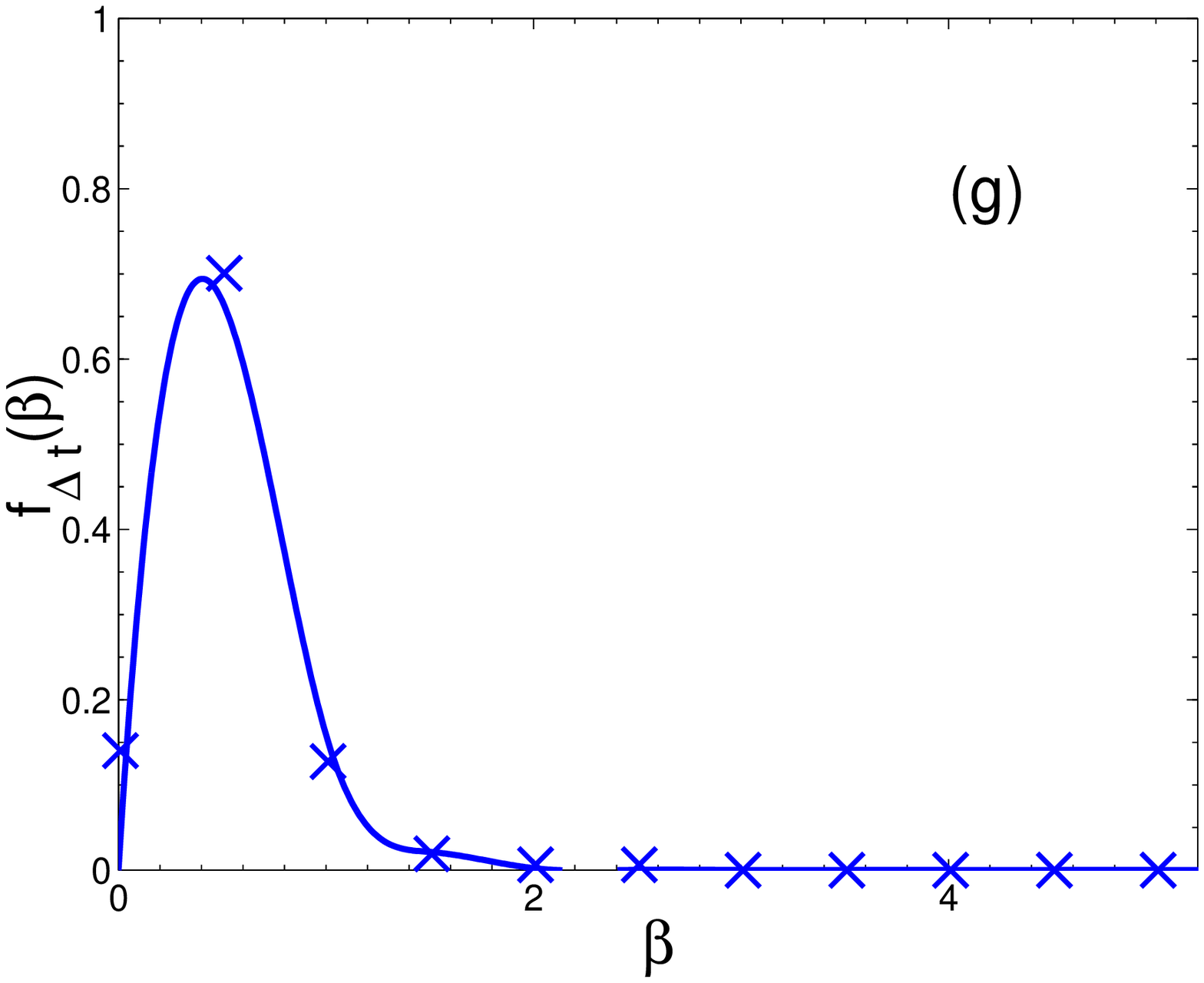} \hfill \leavevmode
\epsfysize=3.7cm \epsffile{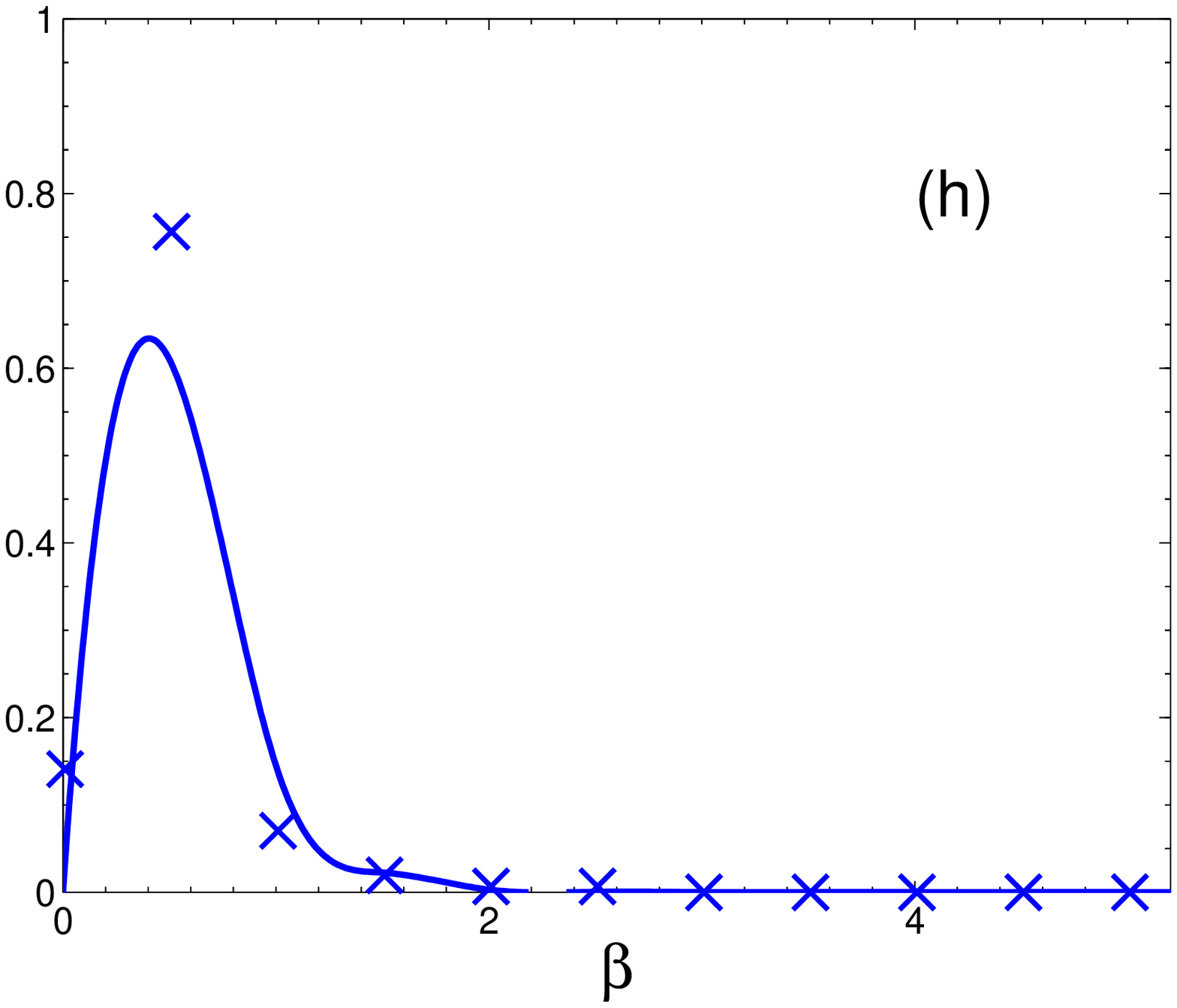} \hfill \leavevmode \epsfysize=3.7cm
\epsffile{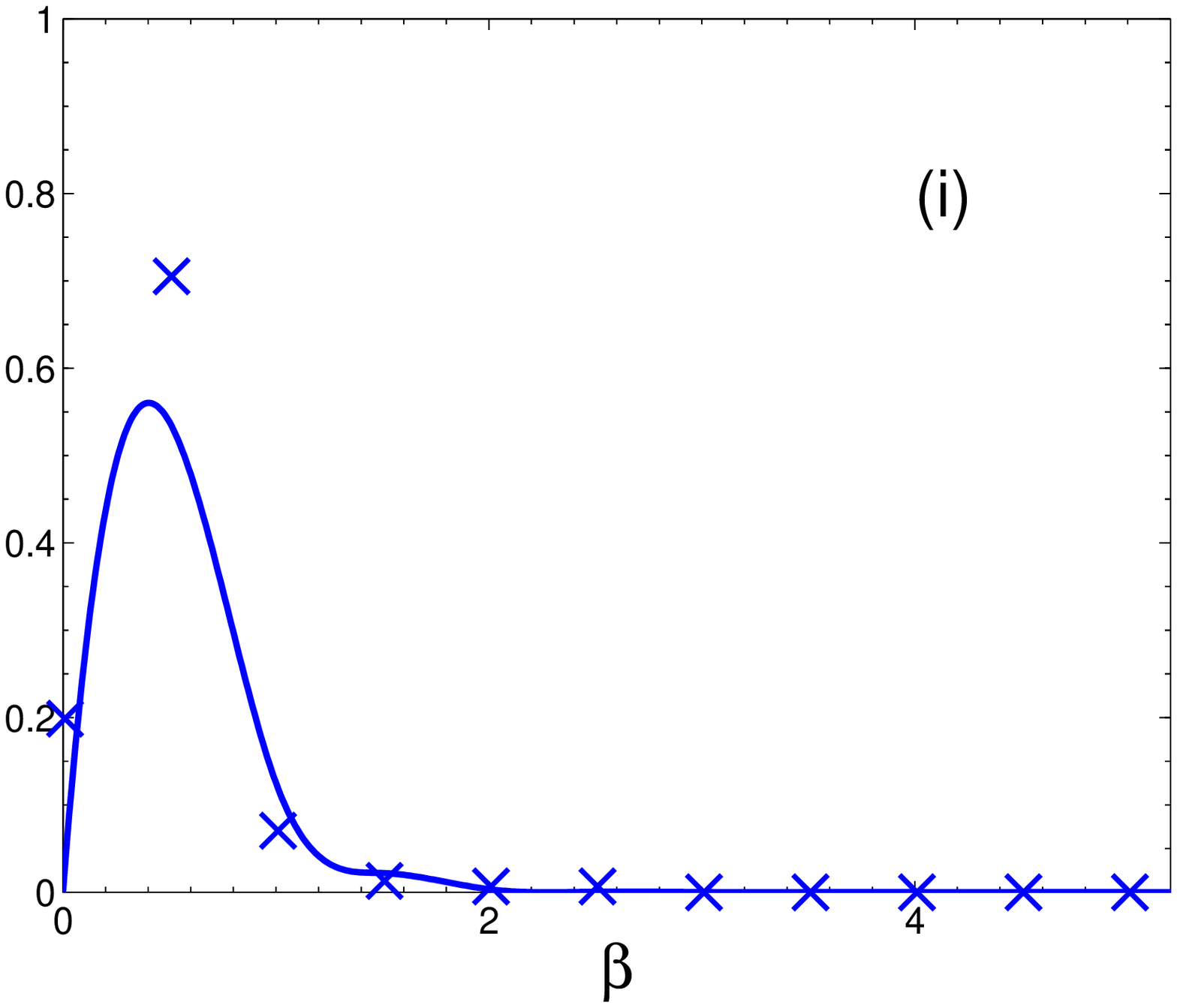} \end{center} \caption{Probability density $f_{\Delta
t}(\beta)$ of the local volatility $\beta$ (Eq.(\ref{volatility})) in terms of
standard deviation of the normalized log returns $Z(t,\Delta t)$ of S\&P500 in
non-overlapping windows with size $m$=32~days for different time lags (symbols)
(a-i) $\Delta t=$1, 5, 10, 15, 20, 25, 30, 35, 40~days. Lines:
$\chi^2$-distribution as given by Eq. (\ref{chi})} \label{fig7}\end{figure}

\newpage \begin{figure}[ht] \begin{center} \leavevmode \epsfysize=5cm
\epsffile{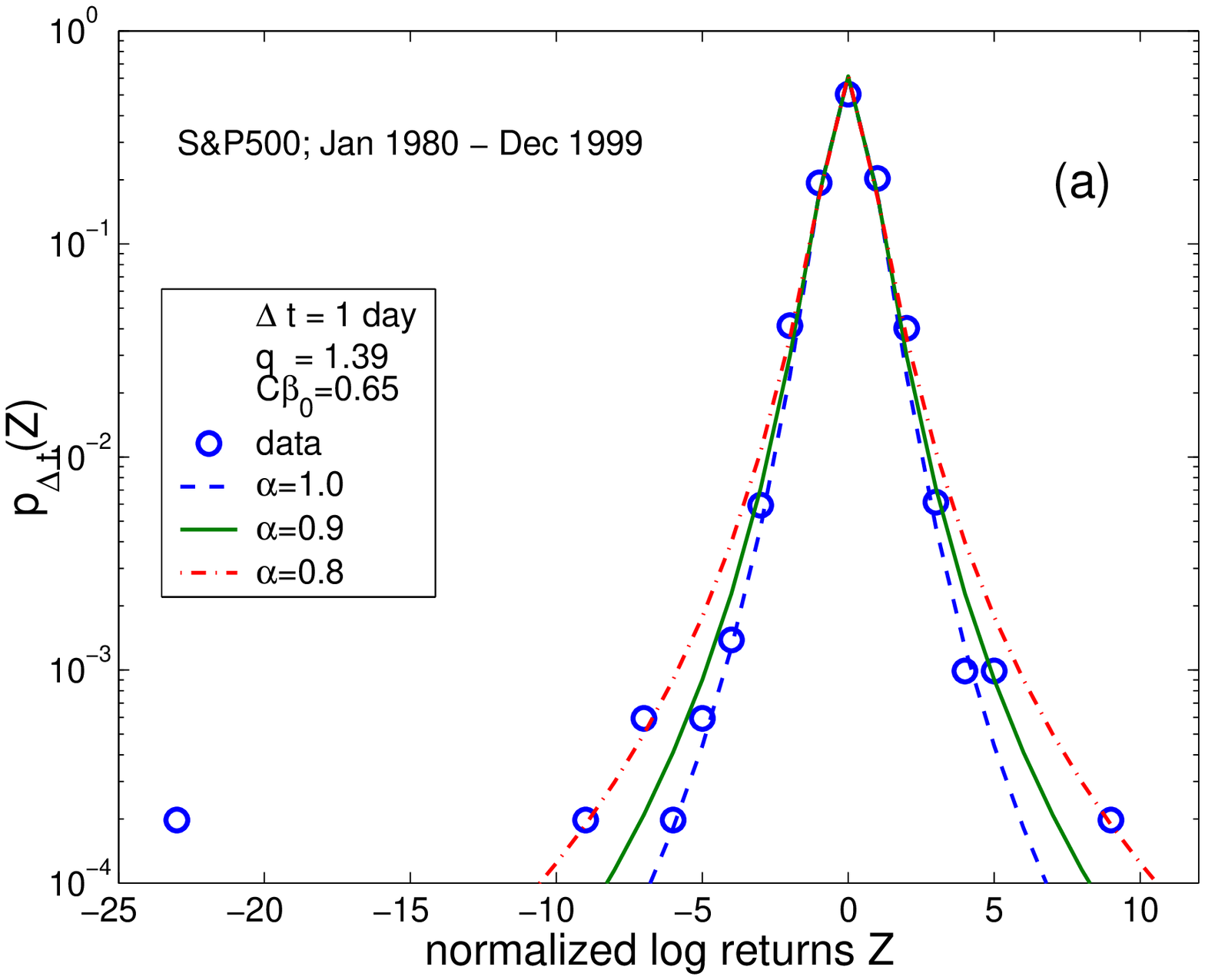} \hfill \leavevmode \epsfysize=5cm \epsffile{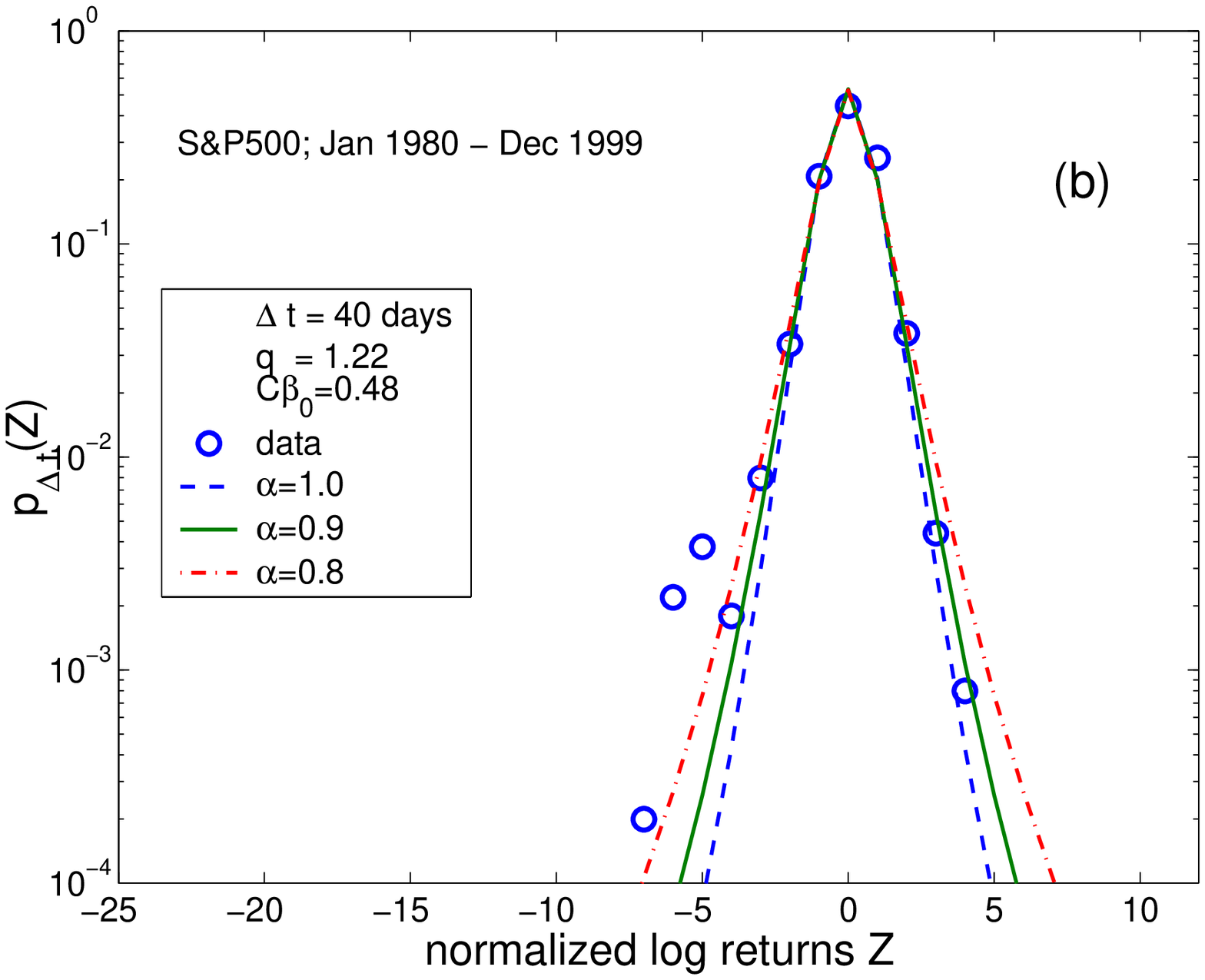}
\vfill \leavevmode \epsfysize=5cm \epsffile{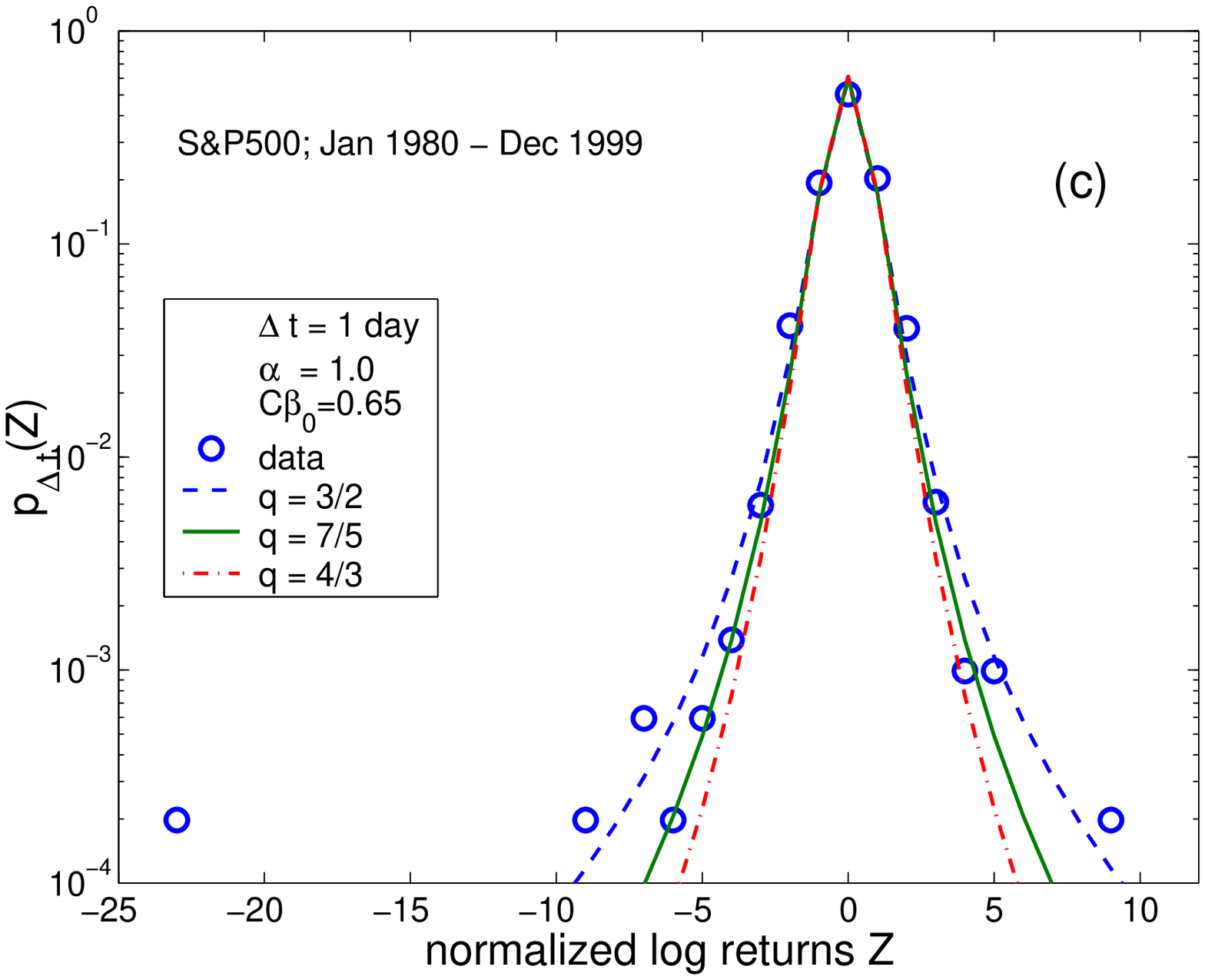} \hfill \leavevmode
\epsfysize=5cm \epsffile{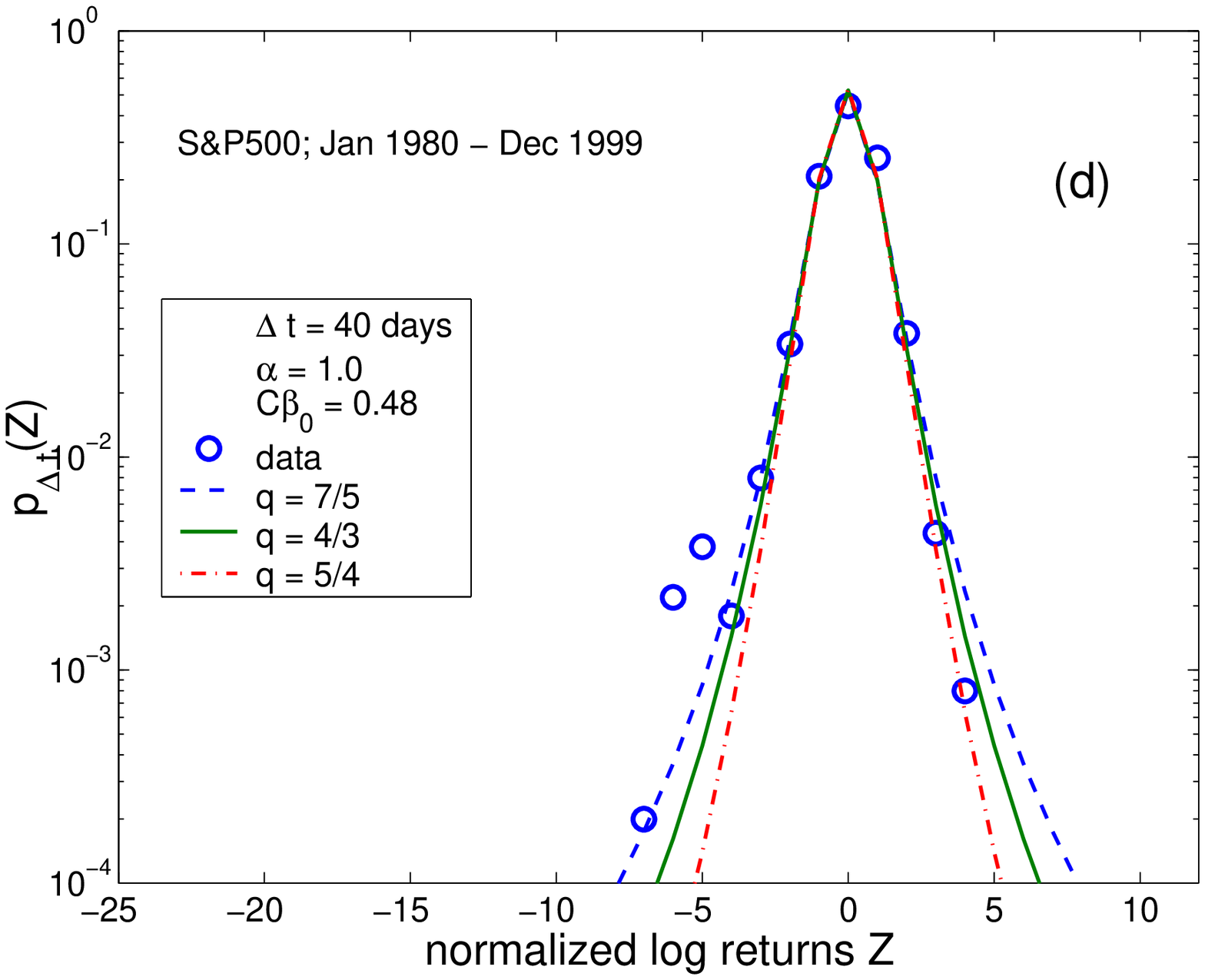} \caption{ Probability 
distribution functions
of the normalized log returns of daily closing price signal of 
S\&P500 (symbols)
for (a) $\Delta t=1$~day and fixed $q=1.39$. The Tsallis type distribution
functions (Eq.(\ref{tsallis})) obtained for various values of the parameter
$\alpha=1.0,0.9,0.8$, dashed, solid, dash-dotted line, respectively; 
(b) same as
(a) but for $\Delta t=40$~days and $q=1.22$; (c) for $\Delta t=1$~day and fixed
$\alpha=1.0$ for various values of $q=3/2, 7/5, 4/3$, dashed, solid, 
dash-dotted
line, respectively; (d) same as (c) but for $\Delta t=40$~days and $q=7/5, 4/3,
5/4$} \end{center}\label{fig8} \end{figure}

\newpage \begin{figure}[ht] \begin{center} \leavevmode \epsfysize=7cm
\epsffile{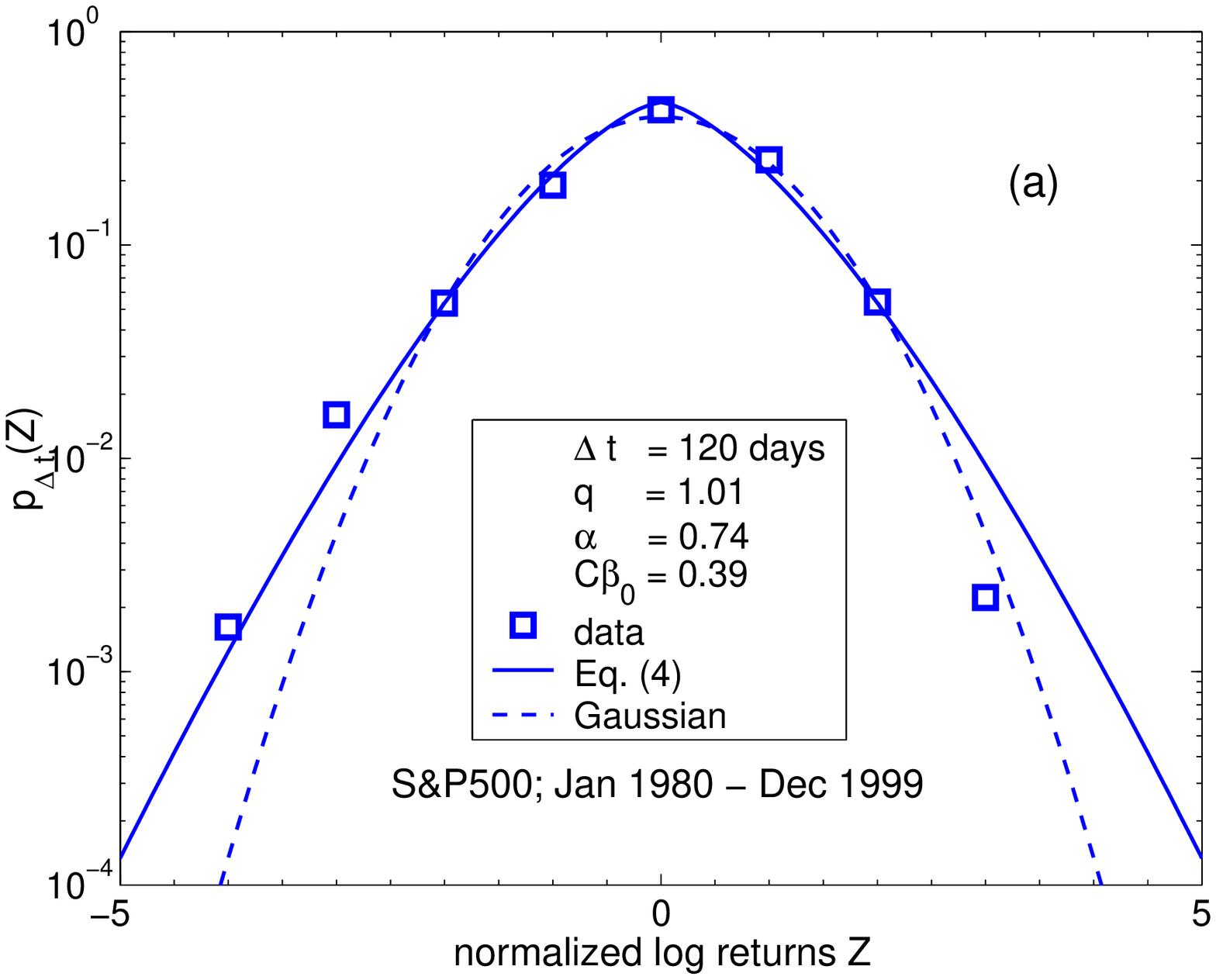} \vfill  \leavevmode \epsfysize=7cm \epsffile{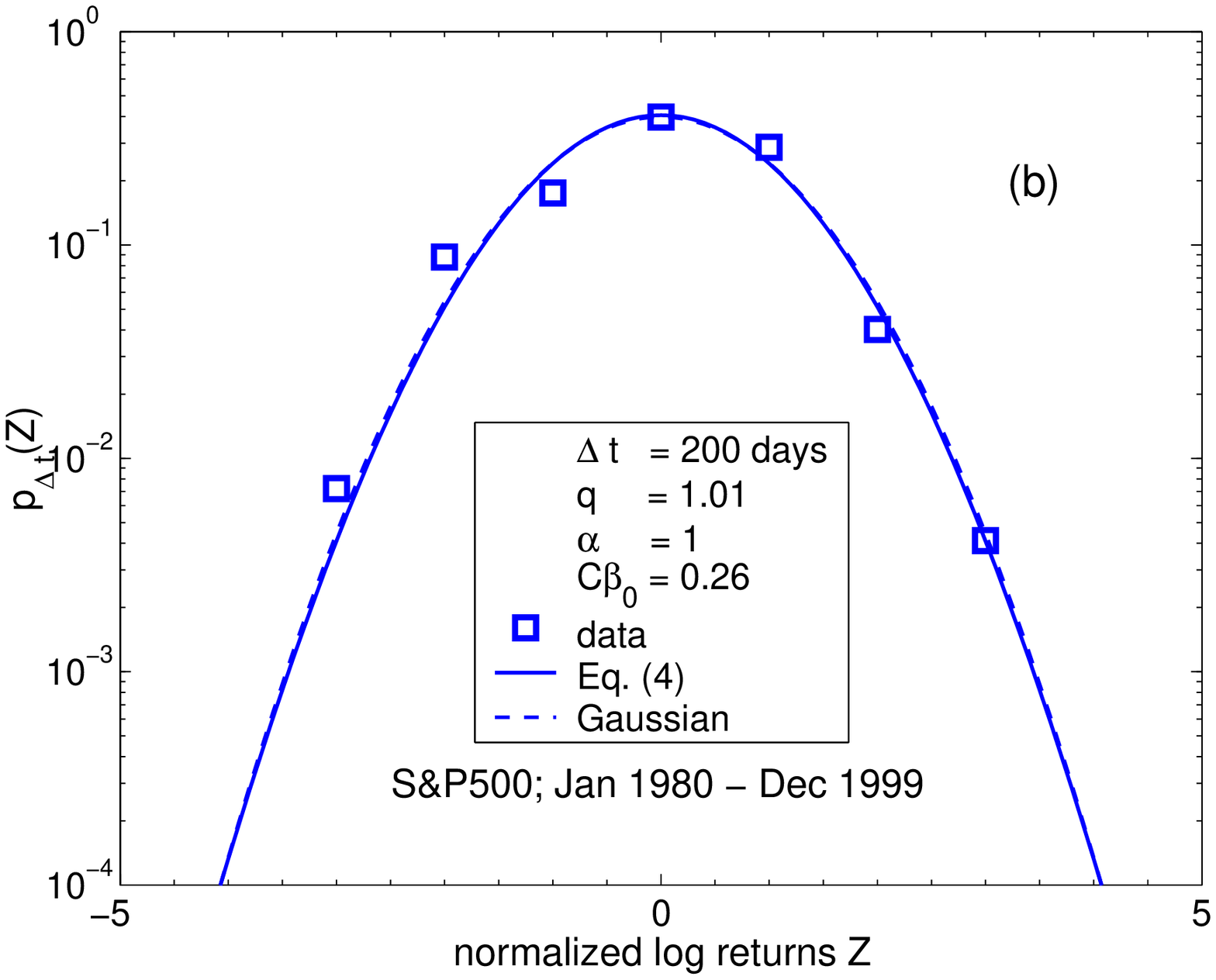}
\caption{Partial distribution function of normalized log return of 
daily closing
price of S\&P500 for a large time lag, i.e. (a) $\Delta t=120$~days and (b)
$\Delta t=200$~days. The solid line marks the best fit with a Tsallis type
distribution function, Eq. (\ref{tsallis}), while the Gaussian distribution
function is drawn with a dashed line} \end{center}\label{fig9} \end{figure}

\newpage \begin{figure}[ht] \begin{center} \leavevmode \epsfysize=10cm
\epsffile{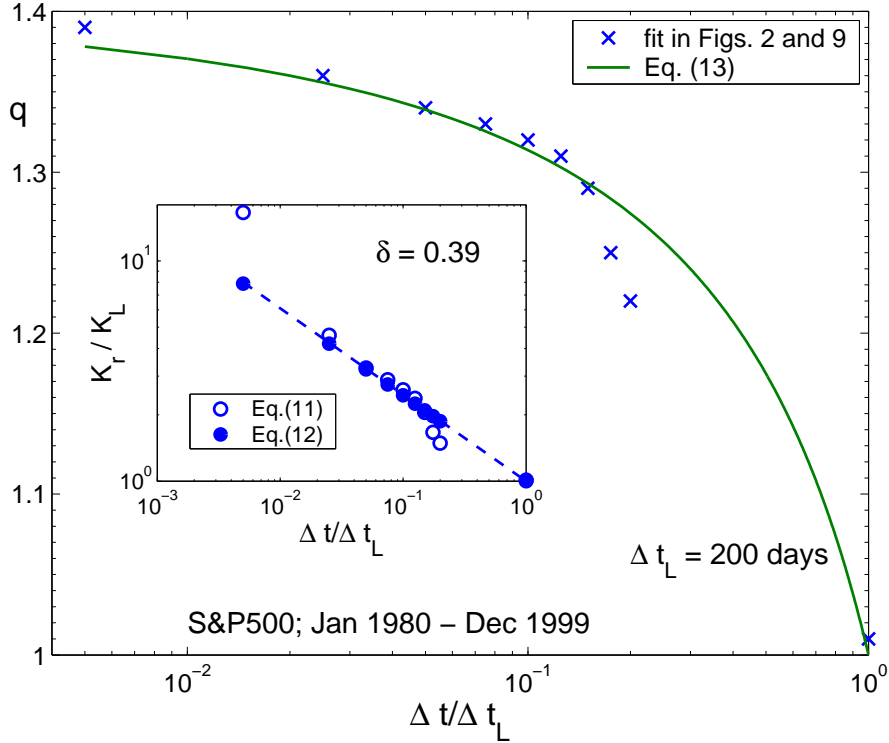} \caption{The functional dependence of the Tsallis $q$
parameter on the rescaled time lag $\Delta t/\Delta t_L$ for $\Delta
t_L=200$~days and $\delta= 0.39$ (see Eq. (\ref{qkr})) (line);  the symbols
represent the values of the $q$ parameter listed in Table \ref{table3} and used
to plot the fitting lines in Figs. 2 and 9. Inset : Scaling properties of the
rescaled kurtosis $K_r/K_L$, where $K_L=3$ is the kurtosis for a Gaussian
process, as a function of the rescaled time lag $\Delta t/\Delta t_L$ 
satisfying
Eq. (\ref{kr}) (open symbols) and Eq. (\ref{krL}) (full symbols)}
\end{center}\label{fig10} \end{figure}

\newpage \begin{figure}[ht] \begin{center} \leavevmode \epsfysize=10cm
\epsffile{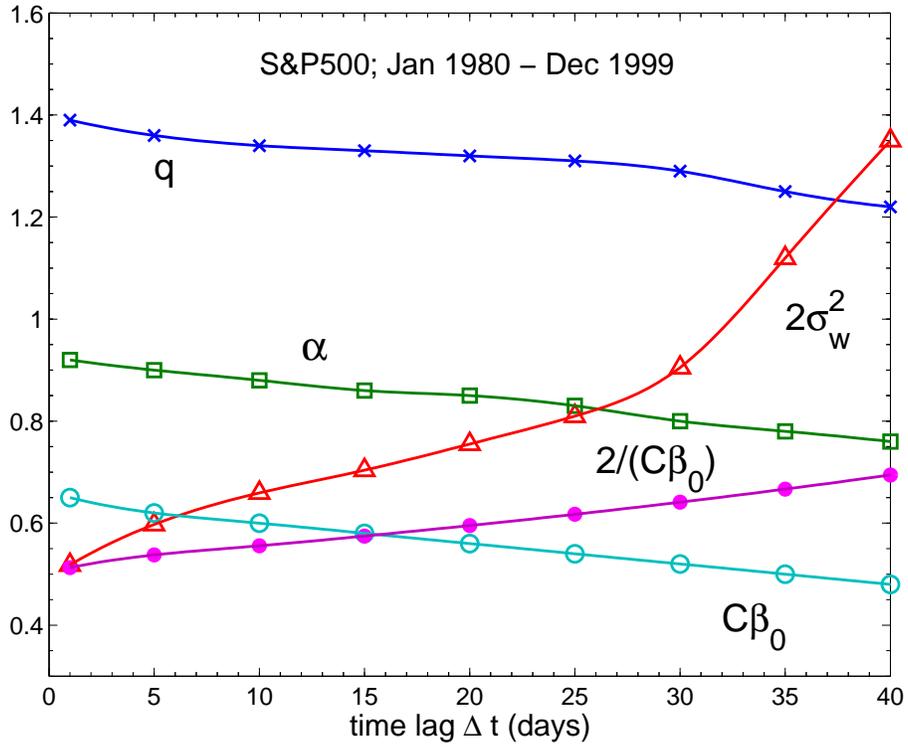} \caption{Characteristic parameters of Tsallis type
distribution function as defined in \cite{kozuki} : Tsallis $q$-parameter
(crosses), $\alpha$ (squares), constant $C\beta_0$ used in the fit (open
circles), the width of the Tsallis type distribution 
$2\sigma_w^2=(2\alpha-(q-1))
/ (2\alpha C\beta_0(q-1))$ from Eq.(\ref{tsallis}) (triangles) (rescaled by a
factor of 1/6), asymptotic behavior of $2\sigma_w^2\approx 2/(C\beta_0)$ for
$\alpha\longrightarrow 1$ (full circles) (rescaled by a factor of 1/6) }
\end{center}\label{fig11} \end{figure}

\begin{figure}[ht] \begin{center} \leavevmode \epsfysize=6.8cm
\epsffile{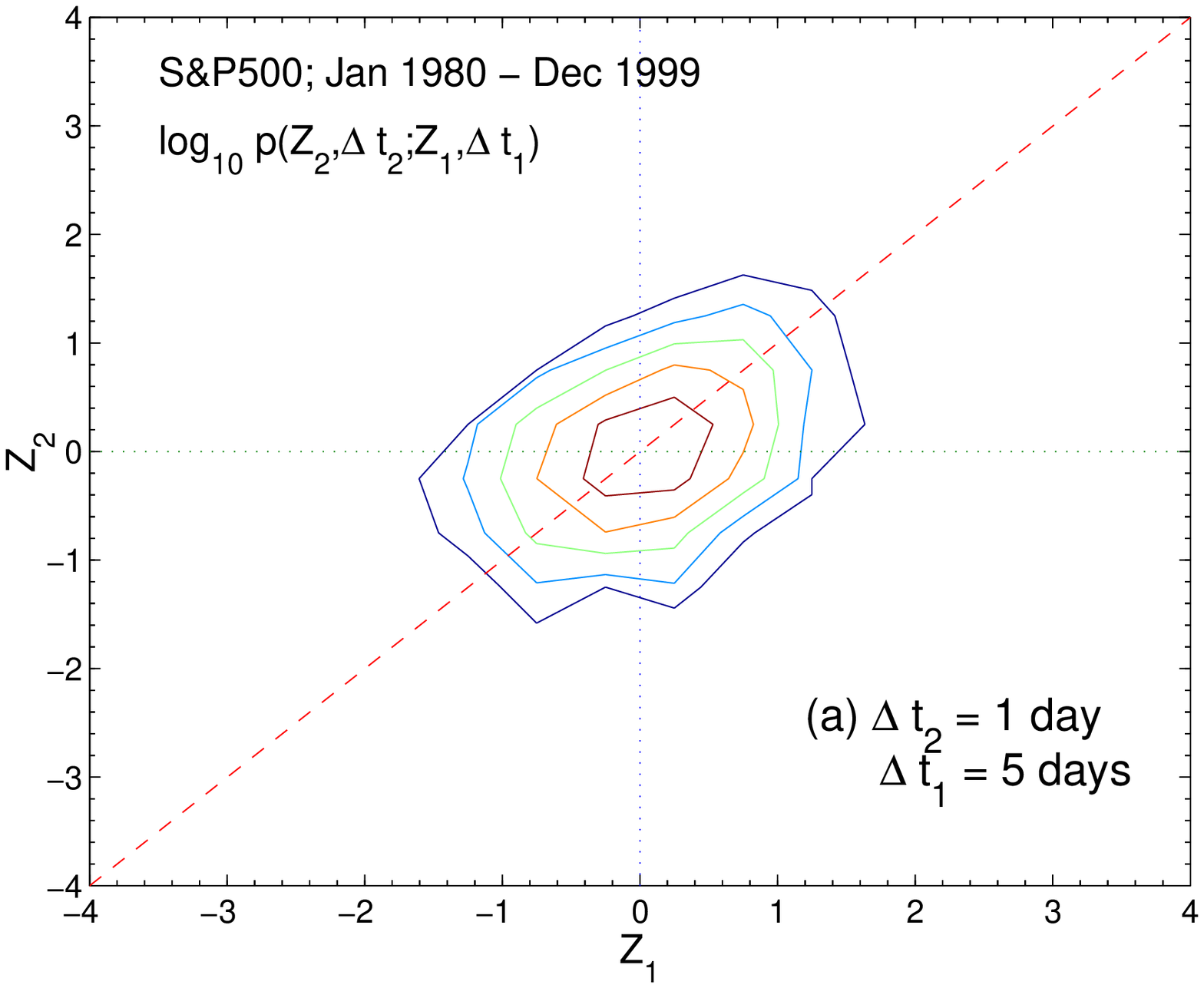} \vfill \leavevmode \epsfysize=6.8cm 
\epsffile{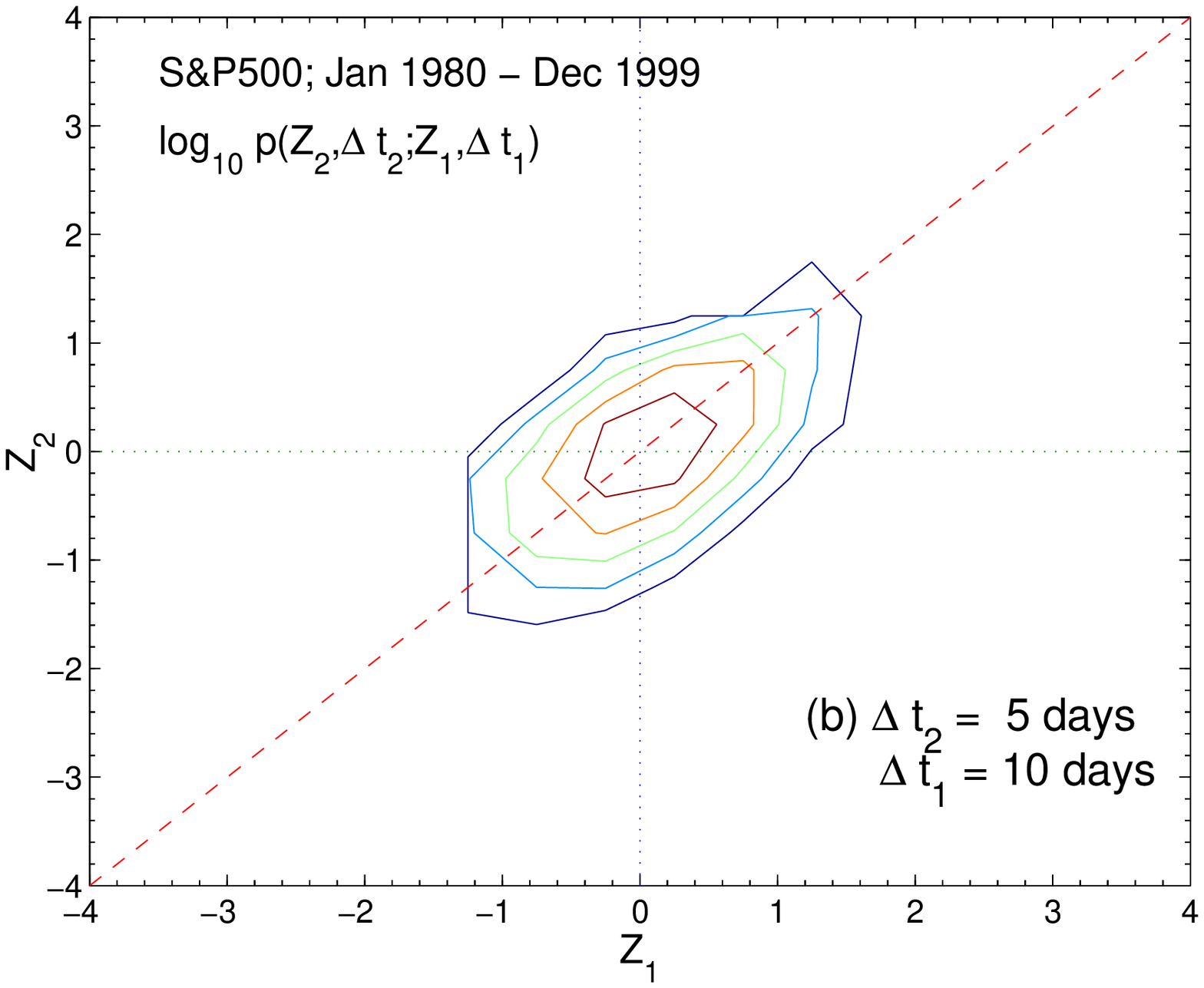}
\end{center} \caption{Typical contour plots of the joint probability density
function $p(Z_2,\Delta t_2; Z_1,\Delta t_1)$ of daily closing price of S\&P500
for the period of interest Jan. 01, 1980 and Dec. 31, 1999. Dashed lines have a
slope +1 and emphasize the correlations between probability density 
functions for
(a) $\Delta t_2=1$~day and $\Delta t_1=5$~days and (b) $\Delta t_2=5$~days and
$\Delta t_1=10$~days. Contour levels correspond to $log_{10}p(Z_2,\Delta t_2;
Z_1,\Delta t_1) =-1.0,-1.5,-2.0,-2.5,-3.0$ from center to border}\label{fig12}
\end{figure}

\begin{figure}[ht] \begin{center} \leavevmode \epsfysize=7cm
\epsffile{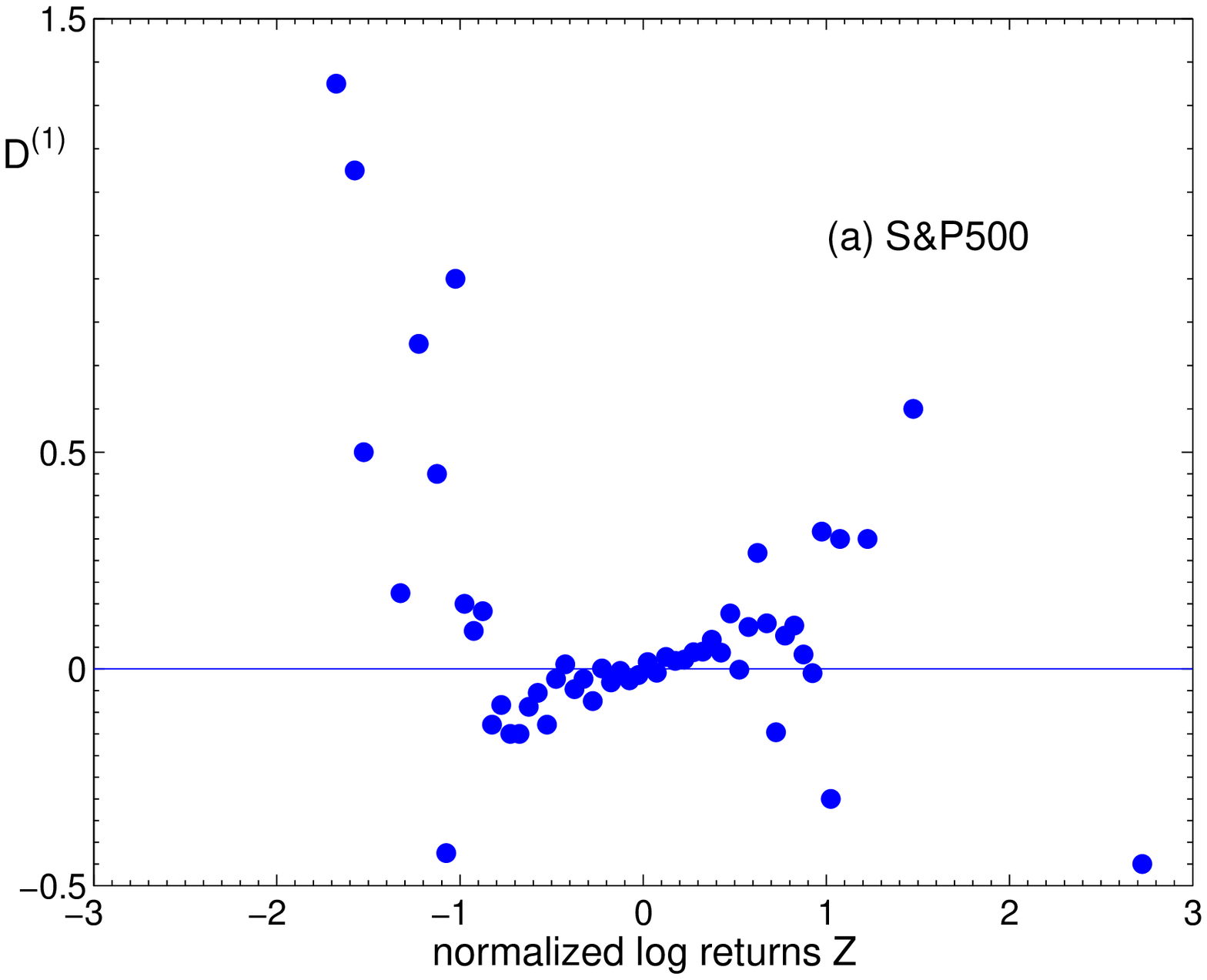} \vfill \leavevmode \epsfysize=7cm \epsffile{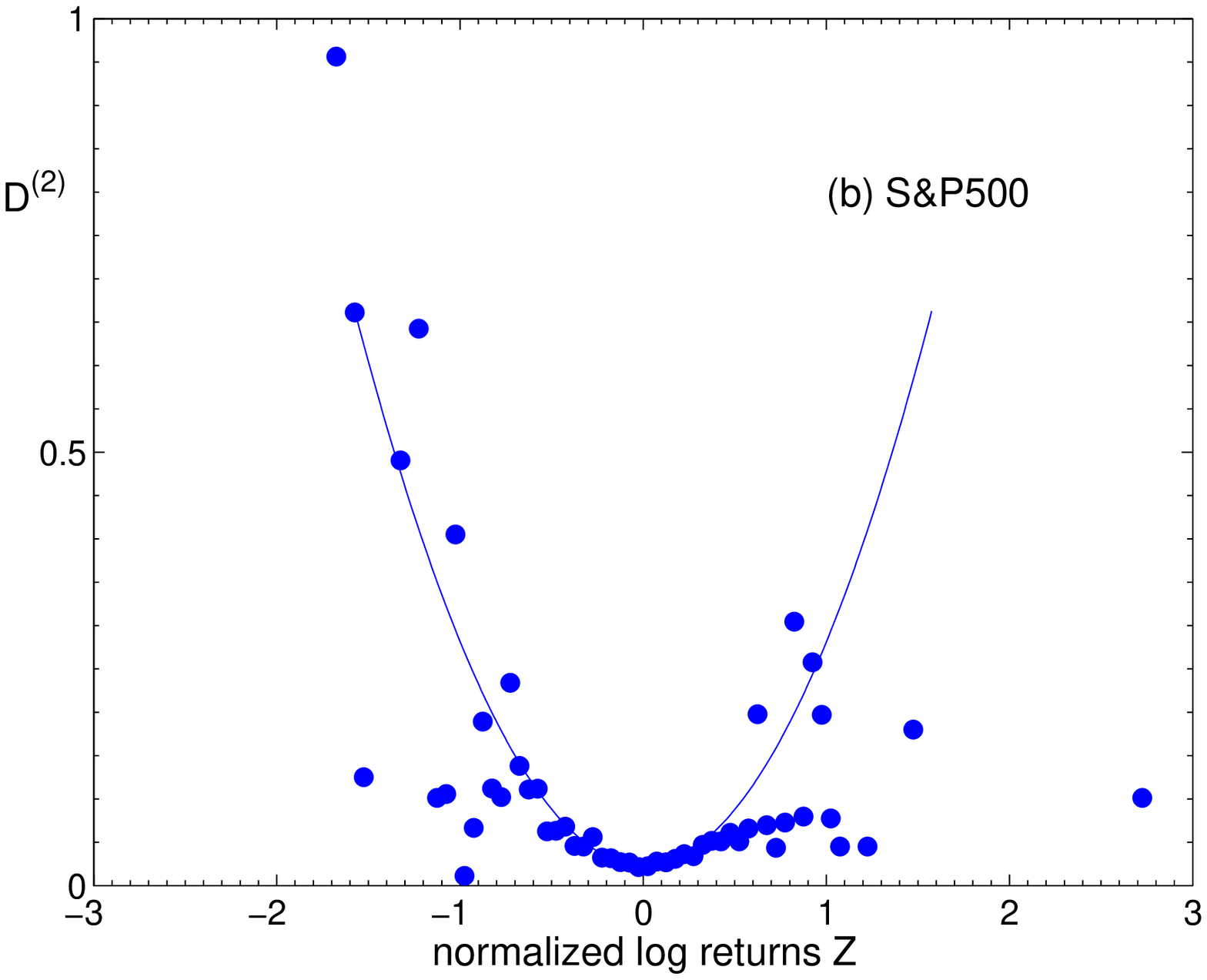}
\end{center} \caption{Kramers-Moyal drift and diffusion coefficients (a)
$D^{(1)}$ and (b) $D^{(2)}$ as a function of normalized log returns 
$Z$ for daily
closing price of S\&P500 ; $ D^{(2)} = 0.26 Z^2 - 0.0005 Z + 0.02$}
\label{fig13}\end{figure}

\begin{figure}[ht] \begin{center} \leavevmode \epsfysize=5cm
\epsffile{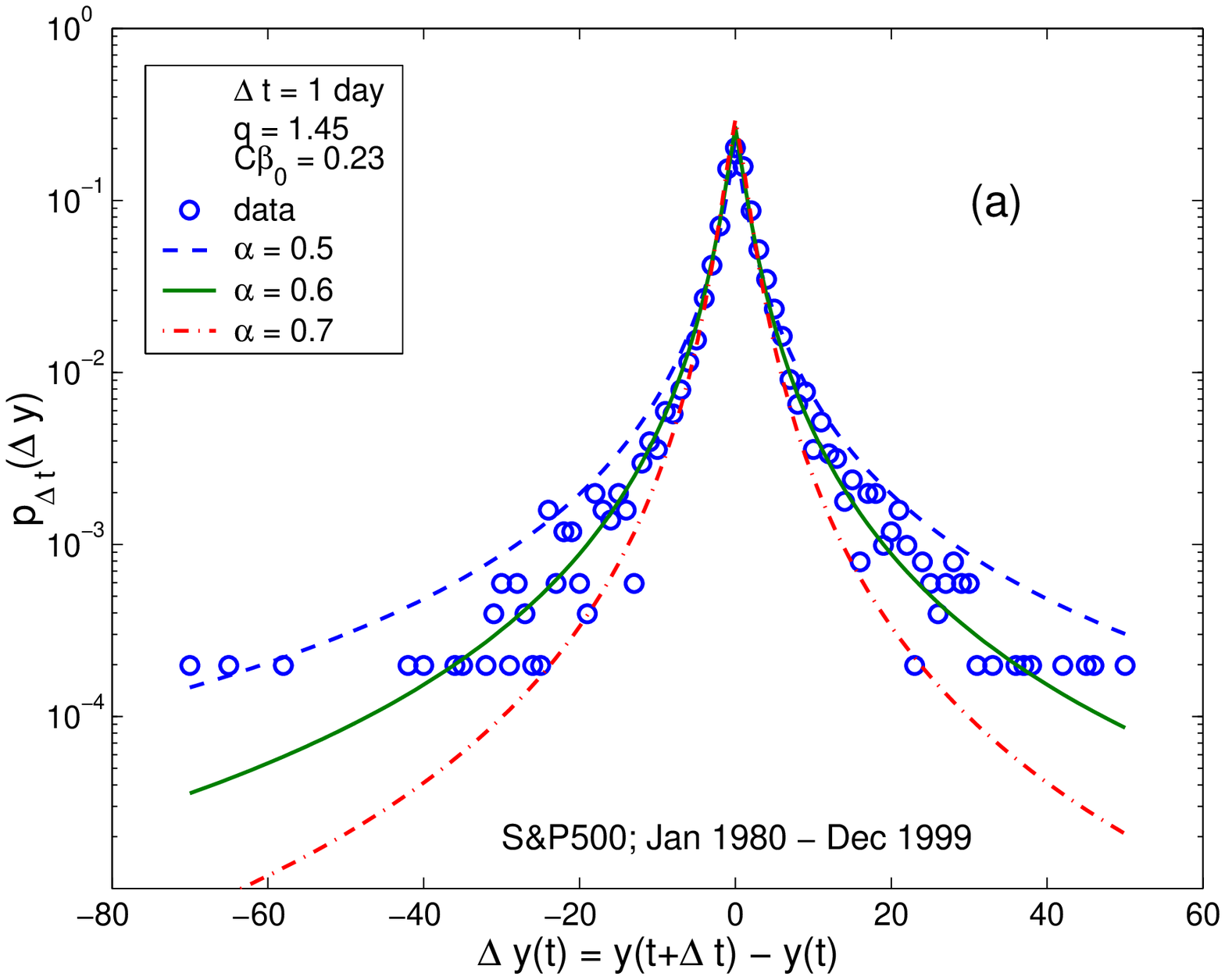} \hfill \leavevmode \epsfysize=5cm \epsffile{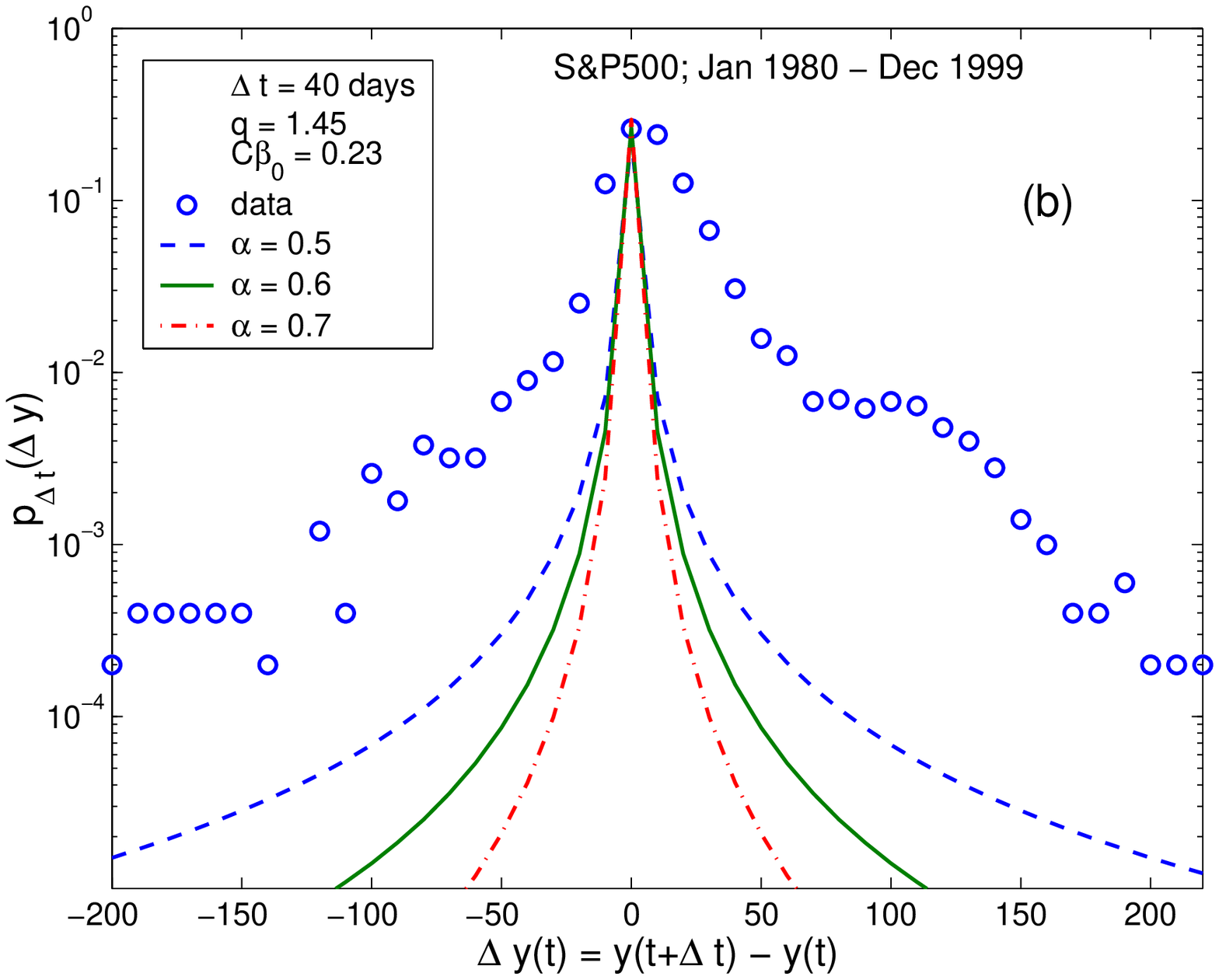}
\vfill \leavevmode \epsfysize=5cm \epsffile{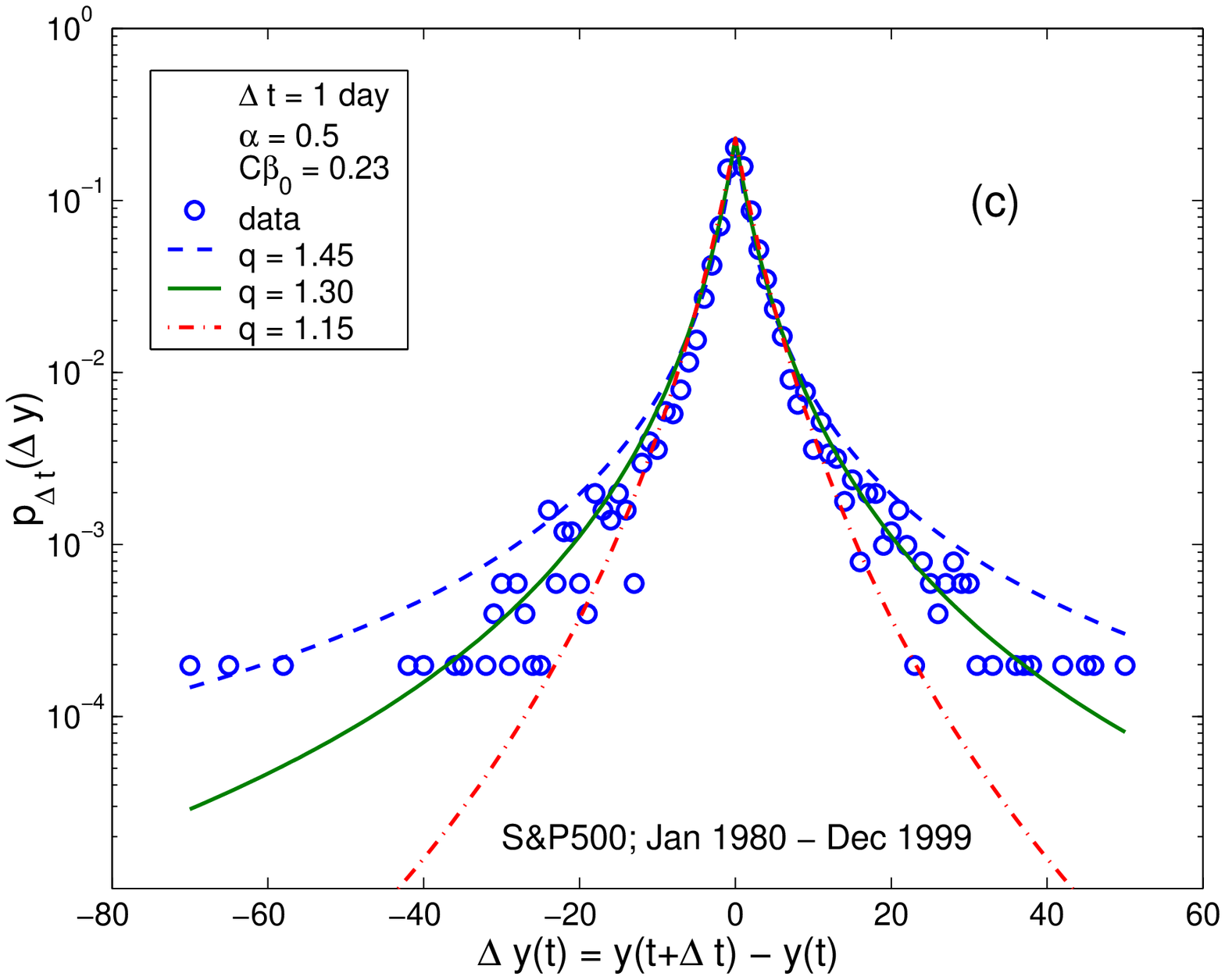} \hfill \leavevmode
\epsfysize=5cm \epsffile{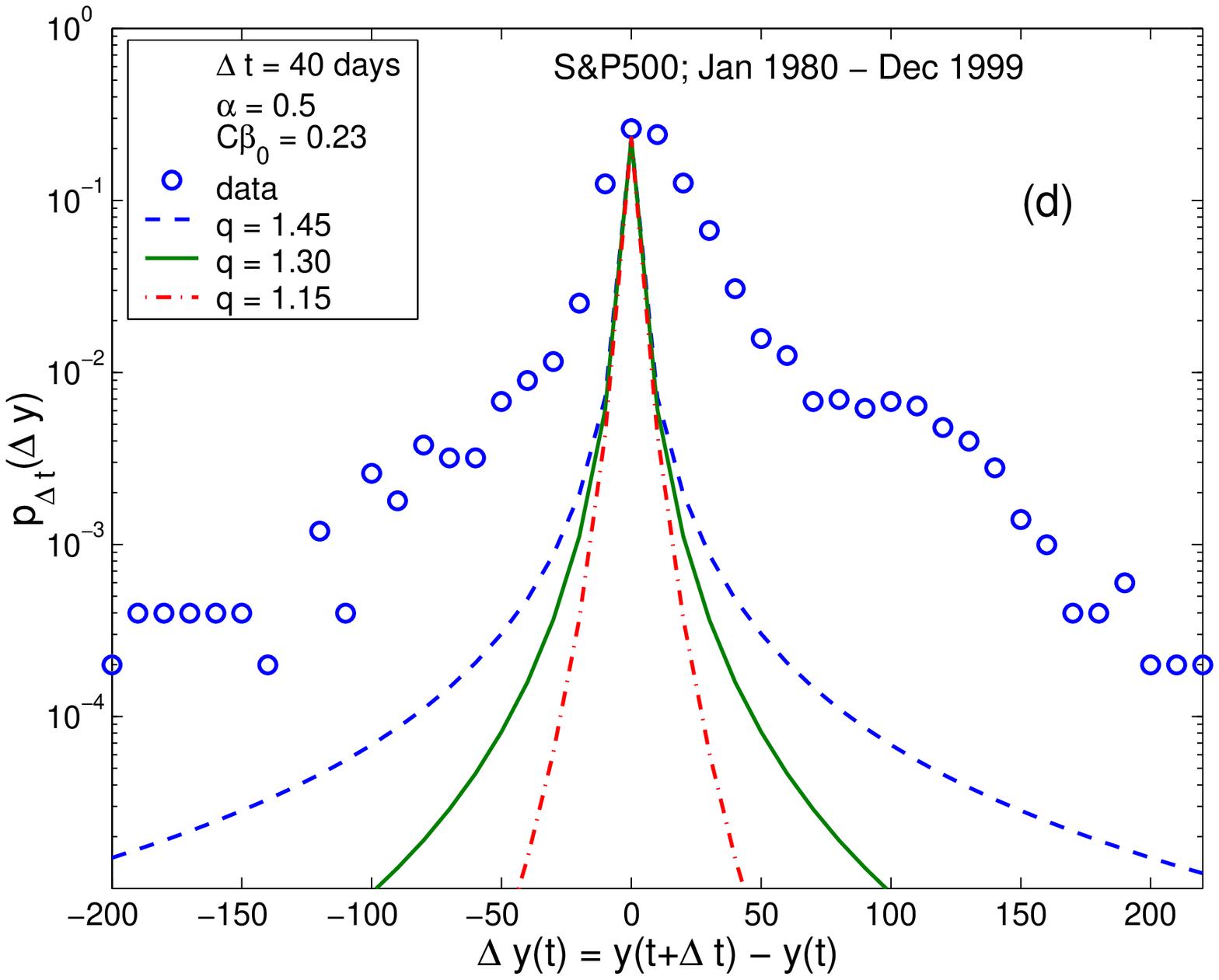} \end{center} \caption{Probability
distribution functions of the daily closing price increments $\Delta
y(t)=y(t+\Delta t)-y(t)$ of S\&P500 (symbols). The Tsallis type distribution
functions (Eq.(\ref{tsallis})) obtained (a) for $\Delta t=1$~day and fixed
$q=1.45$ ($C\beta_0=0.23$) and for various values of the parameter
$\alpha=0.5,0.6,0.7$, dashed, solid, dash-dotted line, respectively; 
(b) the same
as (a) but for $\Delta t=40$~days; (c) for $\Delta t=1$~day and fixed
$\alpha=0.5$ ($C\beta_0=0.23$) and for various values of $q=1.45,1.30,1.15$,
dashed, solid, dash-dotted line, respectively; (d) the same as (b) but for
$\Delta t=40$~days} \label{fig14}\end{figure}

\end{document}